\documentclass[final,Preprint]{zakopane-ab}
 \usepackage{pstricks,pst-node,pst-text,pst-3d}
  \usepackage[dvips,final]{graphicx}
   \usepackage{amsmath}
    \usepackage{amssymb}
     \usepackage{pifont}   
      \usepackage{psfrag}
       \usepackage{bm}

\newrgbcolor{violet}{0.4 0 0.9}                                    %%%%%%%%%
\newcommand{\va}[1]{\langle{#1}\rangle}                            %%%%%%%%%
\newcommand\BraSquare[4]{%                                         %%%%%%%%%
 \begin{picture}(0,0)%                                             %%%%%%%%%
  \setlength{\unitlength}{1pt}%                                    %%%%%%%%%
  \put(#1,#2){\rotatebox{#3}%                                      %%%%%%%%%
              {\raisebox{0mm}[0mm][0mm]{%                          %%%%%%%%%
                \makebox[0mm]{$\left.\rule{0mm}{#4pt}\right]$}}}}  %%%%%%%%%
 \end{picture}}                                                    %%%%%%%%%
\begin{document}
 \title{QCD Sum Rules: From quantum-mechanical oscillator 
        to pion structure in QCD\thanks{E-mail: bakulev@theor.jinr.ru}}
  \author{Alexander~P.~Bakulev
   \address{Bogolyubov Lab. Theor. Phys., JINR (Dubna, Russia)}}
\maketitle

\begin{abstract}
 We illustrate the general scheme of the Sum Rule (SR) method
 using 2D Quantum Harmonic Oscillator (2DQHO) as a toy model.
 We introduce correlator, related to Green function of 2DQHO,
 and describe the property of Asymptotic Freedom for 2DQHO.
 We explain how the duality conception allows one 
 to describe excited states. Finally we present numerical results 
 and extract some lessons to learn from our exposition.
 Then we switch to the QCD and show that QCD SRs supply
 us the method to study  hadrons in non-perturbative QCD.
 Here the main emphasis is put on the pion 
 and its distribution amplitude and form factors.
\end{abstract}
\eqsec
\section{Quantum-mechanical toy model}
Two-dimensional oscillator
with potential ${V(\vec{r}) = m\omega^2r^2/2}$
is the simplest system with confinement.
We select this particular case $D=2$ because all formulas greatly simplify,
for example,
energy levels and wave function values in the origin are
\begin{eqnarray}
 {E_n = (2n+1)\omega\,;}\qquad
 {|\psi_n(0)|^2 = \frac{m\omega}{\pi}\,.}
\end{eqnarray}
We will consider the regular quasi-perturbative method 
of Sum Rules (SR)
to determine energy ${E_0}$ and ${|\psi_0(0)|^2}$ 
of the ground state.
We follow here partially the lectures~\cite{Rad98}. 

The general scheme of the SR method~\cite{SVZ} can be easily understood
on the example of a correlator ${M(\mu)}$,
which has the spectral expansion:
\begin{eqnarray}
 \label{eq:M.spec}
  M^\text{spec}(\mu) 
    =  |\psi_0(0)|^2\, e^{-E_0/\mu}
    + \text{``higher states''}\,.
\end{eqnarray}
Suppose that we can construct the perturbative expansion of this correlator:
\begin{eqnarray}
 \label{eq:M.pert}
  M^\text{pert}(\mu) 
    = M_0(\mu) 
    + \sum_{n\geq1}C_{2n}\frac{\omega^{2n}}{\mu^{2n}}\,,
\end{eqnarray}
where ${M_0(\mu)}$ corresponds to ``free movement''
and has the spectral representation:
\begin{eqnarray}
 \label{eq:M0.spec}
  M_0(\mu)
    = \int_0^{\infty}\!\!\!
      \rho_0(E)\,e^{-E/\mu}\,dE\,.
\end{eqnarray}
The sum rule -- it is simply
\begin{eqnarray}
  M^\text{spec}(\mu) = M^\text{pert}(\mu)\,.
 \nonumber%%\label{eq:SR.spec=pert}
\end{eqnarray}
Usually it appears that higher state contributions can be well
approximated by ``free states'' outside interval $(0,S_0)$.
As a result we have SR in the form:
\begin{eqnarray}
 \label{eq:SR.gen}
  |\psi_0(0)|^2\,e^{-{E_0}/\mu}
    = \int_{0}^{S_0}\!\!
       \rho_0(s)\,e^{-s/\mu}\,ds
    + C_{2}\frac{\omega^{2}}{\mu^{2}}
    + C_{4}\frac{\omega^{4}}{\mu^{4}}
    + \ldots
\end{eqnarray}
\textit{Our aim}: 
to determine ${|\psi_0(0)|^2}$ and ${E_0}$ from this SR 
by calculating spectral density 
${\rho_0(E)}$ of ``free particle''
and coefficients ${C_{2n}}$
by demanding stability of this SR 
in variable ${\mu\in[\mu_\textbf{L},\mu_\textbf{U}]}$
with appropriate value of $S_0$.    

In order to select $M(\mu)$ with these properties 
for our 2DQHO
let us consider 2-time Green function\footnote{%
Due to the fact that $\psi_k(0)=0$ for all states
with $L\geq1$ only $S$-states contribute 
to this sum.}
\begin{eqnarray}
 \label{eq:G.00.xt}
   G(0,0|\vec{x},t)
   = \sum_{k\geq0}\psi^{*}_k(\vec{x})\psi_k(0)e^{-iE_kt}\,
\end{eqnarray}
which is the probability amplitude 
for the transition
${(x=0,t=0)\to (\vec{x},t)}$.
To get ${M(\mu)}$ we put $x=0$, $t=1/{i\mu}$:
\begin{eqnarray}
 \label{eq:G.00.0mu}
  M(\mu) = G(0,0|0,1/{i\mu})  
    = \sum_{k\geq0} \big|\psi_k(0)\big|^2e^{-E_k/\mu}
    = M^\text{spec}(\mu)\,.
  \end{eqnarray}
In our case ${\big|\psi_k(0)\big|^2=m\omega/\pi}$,
so we have the \textit{exact result} for our $M(\mu)$:
\begin{eqnarray}
 \label{eq:M.HO.exact}
 M(\mu) = \frac{m\omega}
               {2\pi\, \sinh\left(\omega/\mu\right)}\,.
\end{eqnarray}   
How fast is convergence of spectral expansion (\ref{eq:G.00.0mu}) for $M(\mu)$?
Let us estimate it for $\mu=\omega$. Exact result (\ref{eq:M.HO.exact}) gives us
$M(\omega) = \left(m\omega/2\pi\right)\cdot0.851$,
whereas numerically
$$ M^\text{spec}(\omega) 
   = \frac{m\omega}{2\pi}\, 
     \left(0.736
         + 0.100
         + 0.013
         + 0.002
         + \ldots 
     \right)\,.
$$
Ground state contributes $86\%$, 
first excitation -- $12\%$, while the second -- only $1.5\%$. 

Now we turn to the perturbative expansion of $M(\mu)$
in powers of $(\omega/\mu)$:
\begin{eqnarray}
 \label{eq:M.HO.pert}
  M^\text{pert}(\mu) 
  = \frac{m\mu}{2\pi}\, 
     \left(1
         - \frac{\omega^2}{6\mu^2} 
         + \frac7{360}\frac{\omega^4}{\mu^4}
         - \frac{31}{15120}\frac{\omega^6}{\mu^6}
         + \ldots 
     \right)\,,
\end{eqnarray}
Here ${m\mu/{2\pi}}$ corresponds to the Green function of free particle:
\begin{eqnarray}
 \label{eq:M.free}
  M^\text{free}(\mu) 
  = \frac{m\mu}{2\pi}\,.
\end{eqnarray}
Numerically at ${\mu=\omega}$ we have
$$ M^\text{pert}(\omega) 
   = \frac{m\omega}{2\pi}\, 
   \left(1 
    - 0.167
    + 0.019
    - 0.002
    + \ldots
   \right)\,.
$$
First correction specifies free result by $17\%$, 
while the second -- by $3\%$.
This perturbative expansion can be rewritten 
\begin{eqnarray}
 \label{eq:M.AF} 
  \frac{\displaystyle M(\mu) - M_0(\mu)}{\displaystyle M_0(\mu)}
   = - \frac{\omega^2}{6\mu^2} 
     + \frac7{360}\frac{\omega^4}{\mu^4}
     - \frac{31}{15120}\frac{\omega^6}{\mu^6}
     + \ldots\,,
\end{eqnarray}
that means \textit{Asymptotic Freedom}:
${\displaystyle M(\mu)}$ behaves like ${\displaystyle M_0(\mu)}$ at large ${\displaystyle\mu\gg\omega}$!
And it is violated 
by power corrections
of the type $\left(\omega^2/\mu^2\right)^n$.

\begin{figure}[ht]
 \centerline{\includegraphics[width=0.49\textwidth]{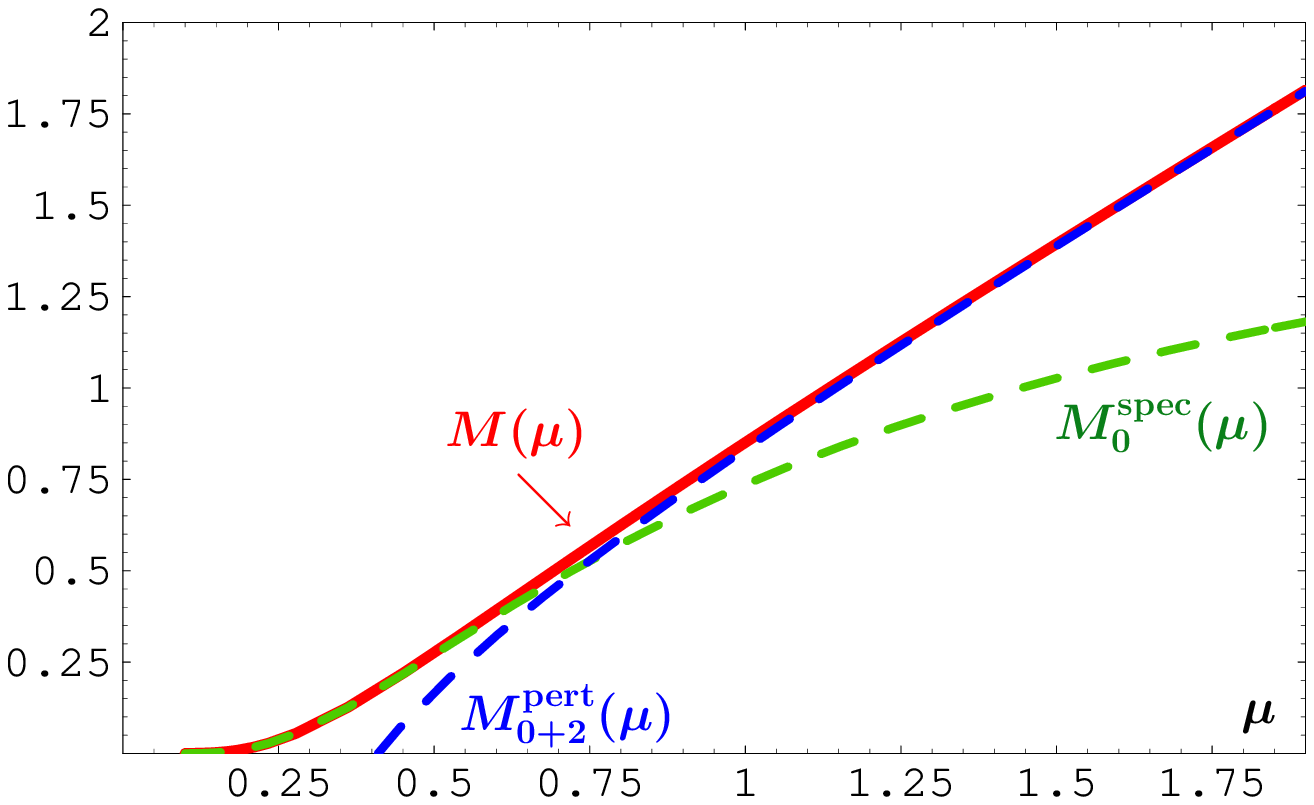}~
             \includegraphics[width=0.49\textwidth]{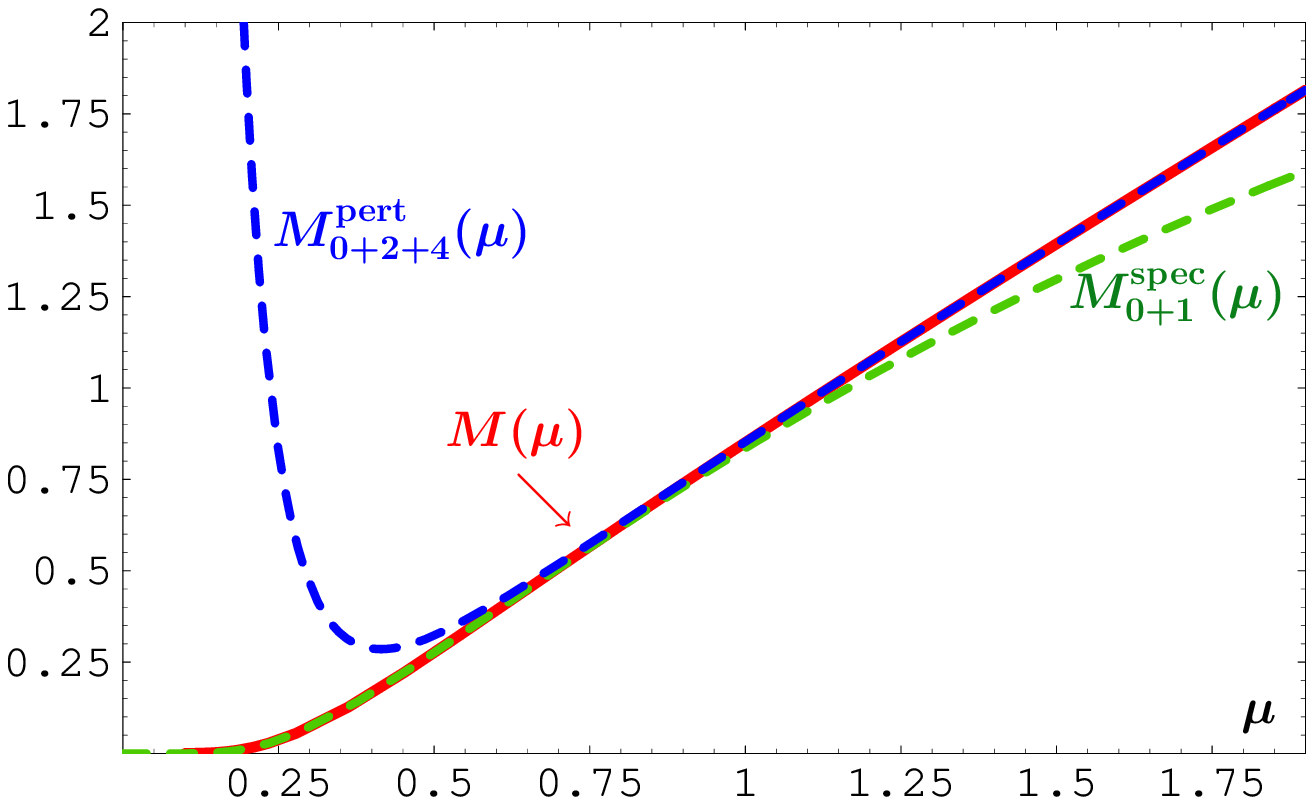}}
  \caption{\footnotesize Left panel: Exact correlator $M(\mu)$, (\ref{eq:M.HO.exact}), is shown 
   by the red solid line, 
   ground state contribution only, $M_0^\text{spec}(\mu)$, -- by the green dashed line,  
   and $M_0(\mu)+O(\omega^2/\mu^2)$ -- by the blue dashed line.
   Right panel: Ground $+$ 1-st excited state contribution only, $M_{0+1}^\text{spec}(\mu)$,
   is shown by the green dashed line, whereas $M_0(\mu)+O(\omega^4/\mu^4)$ 
   -- by the blue dashed line.
   \label{fig:Ex.GrSt.Pert2}}
\end{figure}

\begin{figure}[ht]
 \centerline{\includegraphics[width=0.49\textwidth]{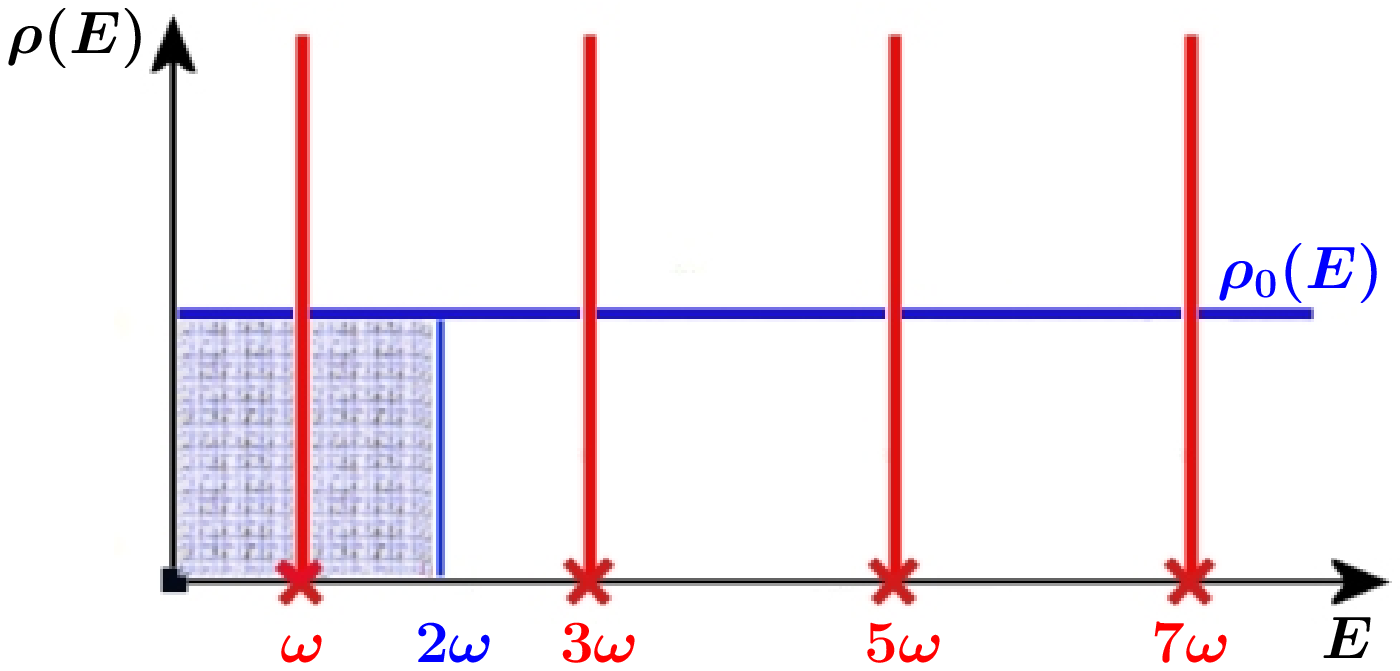}~%
             \includegraphics[width=0.49\textwidth]{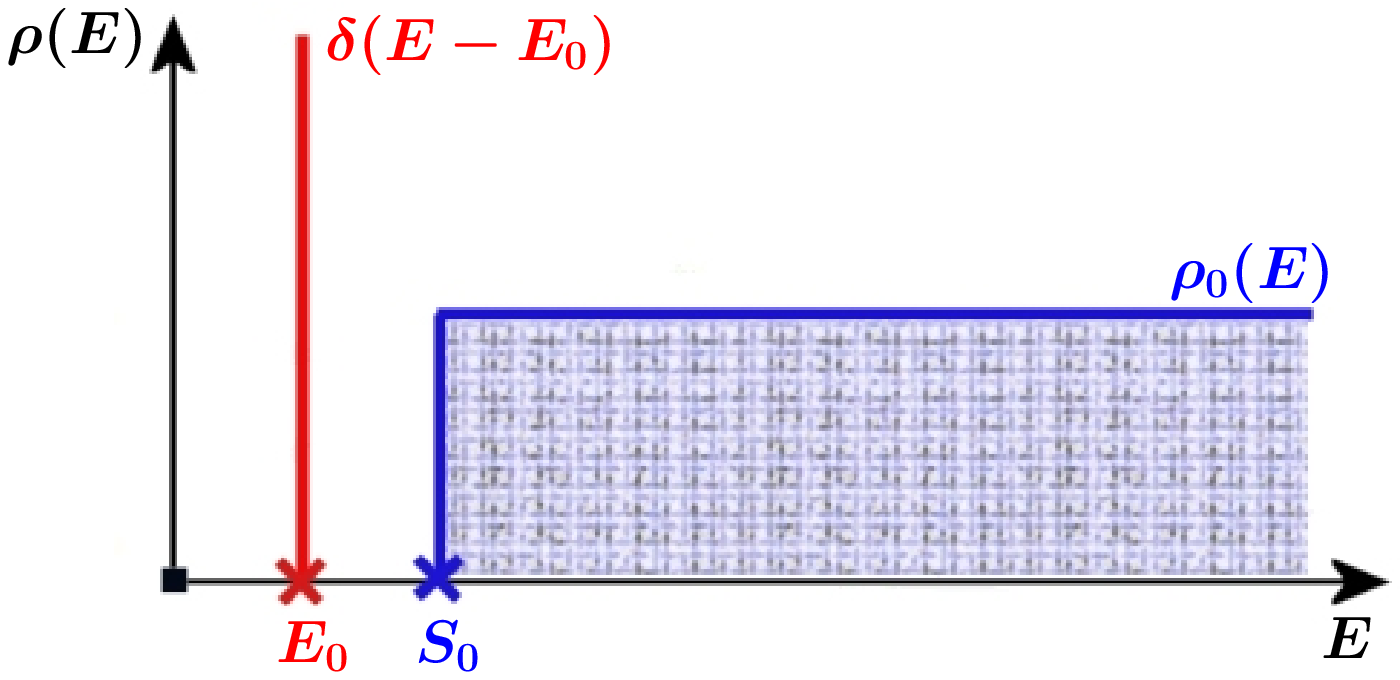}}
  \caption{\footnotesize \textbf{Left panel:} The exact spectral density $\rho^\text{osc}(E)$, 
  (\ref{eq:Sp.Den.Osc}), is shown by red solid vertical lines, 
  imitating $\delta$-functions, 
  whereas free spectral density, $\rho_0(E)$, -- by the blue solid line.
  \textbf{Right panel:} The phenomenological model $\rho^\text{mod}(E)$, 
  (\ref{eq:Sp.Den.Model}), is shown by red solid vertical line, 
  imitating $\delta$-function, and by the blue solid line,
  corresponding to $\rho_0(E)$ and starting from the threshold $S_0$.  
  \label{fig:Dual.Sp.Dens}}
\end{figure}

We see in Fig.\ \ref{fig:Ex.GrSt.Pert2}
%%%%%%%%%%%%%%%%%%%%%%%%%%%%%%%%%%%%%%%%%%%%%%%%%%%%%%%%%%%%%%%%%%%%%%%%%%%%%%%%
%%%%%%%%%%%%%%%%%%%%%%%%%%%%%%%%%%%%%%%%%%%%%%%%%%%%%%%%%%%%%%%%%%%%%%%%%%%%%%%%
that for large $\mu$ asymptotic freedom works well: 
 $M(\mu)\simeq M_0(\mu)$,
but we need more and more resonances to saturate 
 $\displaystyle M(\mu)\vphantom{|\psi_0|^2e^{-E_0/\mu}}$.
For small $\mu$ in spectral part survives only ground state 
$\displaystyle|\psi_0|^2e^{-E_0/\mu}$, 
but the perturbation expansion breaks down.
%%\textit{Global and local dualities}.
In order to model higher resonances 
let us consider the spectral representation 
of our correlator $M(\mu)$:
\begin{eqnarray}
 M^\text{spec}(\mu)
  = \sum_{k\geq0} \frac{m\omega}{\pi}\,e^{-E_k/\mu}
   \equiv \int_0^{\infty}\!\!\!
           \rho^\text{osc}(E)\,e^{-E/\mu}\,dE\,.
\end{eqnarray}
Here the spectral density is just the sum of $\delta$-functions:
\begin{eqnarray}
 \label{eq:Sp.Den.Osc}
 \rho^\text{osc}(E)
   = \sum_{k\geq0} \frac{m\omega}{\pi}\,\delta(E-E_k)\,.
\end{eqnarray}
Analogously we have integral representation for free correlator:
\begin{eqnarray}
 M_0(\mu)
  = \frac{m\mu}{2\pi}
   \equiv \int_0^{\infty}\!\!\!\rho_0(E)\,e^{-E/\mu}\,dE\,,
\end{eqnarray}
where ${\displaystyle\rho_0(E) = \frac{m}{2\pi}}$.
Asymptotic freedom dictates \textit{global duality} of
these two densities
(term ``global'' is related with integration over the whole spectrum)
 \begin{eqnarray}
  \int_0^{\infty}\!\!\!\rho^\text{osc}(E)\,dE
  = \int_0^{\infty}\!\!\!\rho_0(E)\,dE\,.
 \end{eqnarray}  
At first glance these spectral densities have completely different behaviour,
see Fig.\ \ref{fig:Dual.Sp.Dens}.
%%%%%%%%%%%%%%%%%%%%%%%%%%%%%%%%%%%%%%%%%%%%%%%%%%%%%%%%%%%%%%%%%%%%%%%%%%%%%%%%
%%%%%%%%%%%%%%%%%%%%%%%%%%%%%%%%%%%%%%%%%%%%%%%%%%%%%%%%%%%%%%%%%%%%%%%%%%%%%%%%
But we have very interesting relations between 
${2k\omega}$-partial integral moments  of this dual densities,
namely, 
$\langle{E^N}\rangle_{2k\omega}
  \equiv\!\int_{2k\omega}^{2k\omega+2\omega}\!\!E^N\!\rho(E)\,dE$:
\begin{eqnarray}
 \int_{2k\omega}^{2(k+1)\omega}\!\!\!\rho^\text{osc}(E)\,dE\,
         = &\displaystyle\,\frac{m\omega\vphantom{\omega^2}}{\pi}\,
           &
         =\,\int_{2k\omega}^{2(k+1)\omega}\!\!\!\rho_0(E)\,dE\,;~~~\\
 \int_{2k\omega}^{2(k+1)\omega}\!\!\!E\,\rho^\text{osc}(E)\,dE\,
         = &\displaystyle\,\frac{m\omega^2(2k+1)}{\pi}
           & 
         =\,\int_{2k\omega}^{2(k+1)\omega}\!\!\!E\,\rho_0(E)\,dE\,.~~~
\end{eqnarray}
For $N\geq2$ we have approximate relation
$\langle{E^N}\rangle_{2k\omega}^\text{osc}=\langle{E^N}\rangle_{2k\omega}^0%
 \left[1+O\left(N^2/k^2\right)\right]$.
So, we have duality between each excited resonance in oscillator
and free particle in some spectral domain.
That means \textit{local duality}.

Now we can model higher state contributions by
``higher states'' = ``free states'' outside interval $(0,S_0)$
or:
\begin{eqnarray}
 \label{eq:Sp.Den.Model}
  \rho^\text{mod}(E)
     = |\psi_0(0)|^2\,\delta\left(E-E_0\right)
     +  \rho_0(E)\,\theta\left(E-S_0\right)
\end{eqnarray}
and this gives us
\begin{eqnarray}
 M^\text{mod}(\mu)
     = |\psi_0(0)|^2\,e^{-E_0/\mu} 
     + \int_{S_0}^{\infty}\!\!\rho_0(s)\,e^{-E/\mu}\,dE\,.
\end{eqnarray}
After all we have the following SR:
\begin{eqnarray}
 |\psi_0(0)|^2 e^{-{E_0}/\mu}
  = \int_{0}^{S_0}\!\!
     \rho_0(E)\,e^{-E/\mu}\,dE 
    + \text{power corrections}\,,
\end{eqnarray}
or, equivalently,
with ${\Psi_0(0)\equiv\psi_0(0)\sqrt{\pi/\omega}}$:
\begin{eqnarray}
 \label{eq:SR1}
 |\Psi_0(0)|^2 e^{-E_0/\mu}
   = \frac{\mu}{2\omega}
      \left\{1-e^{-S_0/\mu} 
            -\frac{\omega^2}{6\mu^2}
            + \ldots
      \right\}\,.
\end{eqnarray}
We also have a daughter SR
produced from (\ref{eq:SR1})
by applying $\partial/{\partial \mu^{-1}}$:
\begin{eqnarray}
 \label{eq:SR2}
  |\Psi_0(0)|^2E_0\,e^{-E_0/\mu}
    = \frac{\mu^2}{2\omega}
       \left\{1-\left(1+\frac{S_0}{\mu}\right)e^{-S_0/\mu} 
            +\frac{\omega^2}{6\mu^2}
            + \ldots
       \right\}\,.
\end{eqnarray}
If we divide (\ref{eq:SR2}) by (\ref{eq:SR1}) we obtain
SR for $E_0$
\begin{eqnarray}
 \label{eq:SR.E0}
   E_0
    = \mu\,\frac{1-\left(1+S_0/\mu\right)e^{-S_0/\mu} 
                      +\omega^2/(6\mu^2) + \ldots}
                {1-e^{-S_0/\mu}
                  -\omega^2/(6\mu^2) + \ldots}\,.
\end{eqnarray}
The strategy of processing these SRs is:
\begin{itemize}
  \item To determine $E_0 \approx E_0(S_0,\mu)$
  by minimal sensitivity to variation 
  of $\mu\in[\mu_\text{L};\mu_\text{U}]$ 
  at appropriate ${S_0}$;
  \item To determine $|\Psi_0(0)|^2\approx\Psi_0^2(S_0,E_0,\mu)$
  by minimal sensitivity to variation of $\mu$
  at appropriate $S_0$.
\end{itemize}
How we should determine the \textit{fidelity window} $[\mu_\text{L};\mu_\text{U}]$?
Power corrections are of the type ${(\omega/\mu)^{2n}}$ and 
they are huge at ${\mu\ll\omega}$.
We demand: 
\begin{eqnarray}
 \label{eq:Delta.Pert.33}
  \Delta_\text{pert}(\mu)
          \equiv
           \sum_{n\geq1}\frac{\displaystyle C_{2n}(\omega/\mu)^{2n}}
                       {\displaystyle M_0(\mu)}
          \leq 0.33 
          \text{\black~~for all~~}\mu\geq\mu_\text{L}\,.
\end{eqnarray}
Higher states are not suppressed
by ${e^{-E_k/\mu}\approx 1}$
at large ${\mu\gg\omega}$.
We demand: 
\begin{eqnarray}
 \label{eq:Delta.HS.33} 
  \Delta_\text{H.S.}(\mu)
          \equiv
           \int_{S_0}^{\infty}\frac{\rho_0(E)}{M_0(\mu)}\,
            e^{-E/\mu}dE 
          \leq 0.33
           \text{\black~~for all~~}\mu\leq\mu_\text{U}\,.
\end{eqnarray}  \newpage \noindent 
Then the fidelity window is ${\mu_\text{L}\leq\mu\leq\mu_\text{U}}$:
Only for ${\mu}$ inside it is reasonable to demand 
minimal sensitivity of SRs to variations in ${\mu}$!

\begin{figure}[t]
 \centerline{\includegraphics[width=0.49\textwidth]{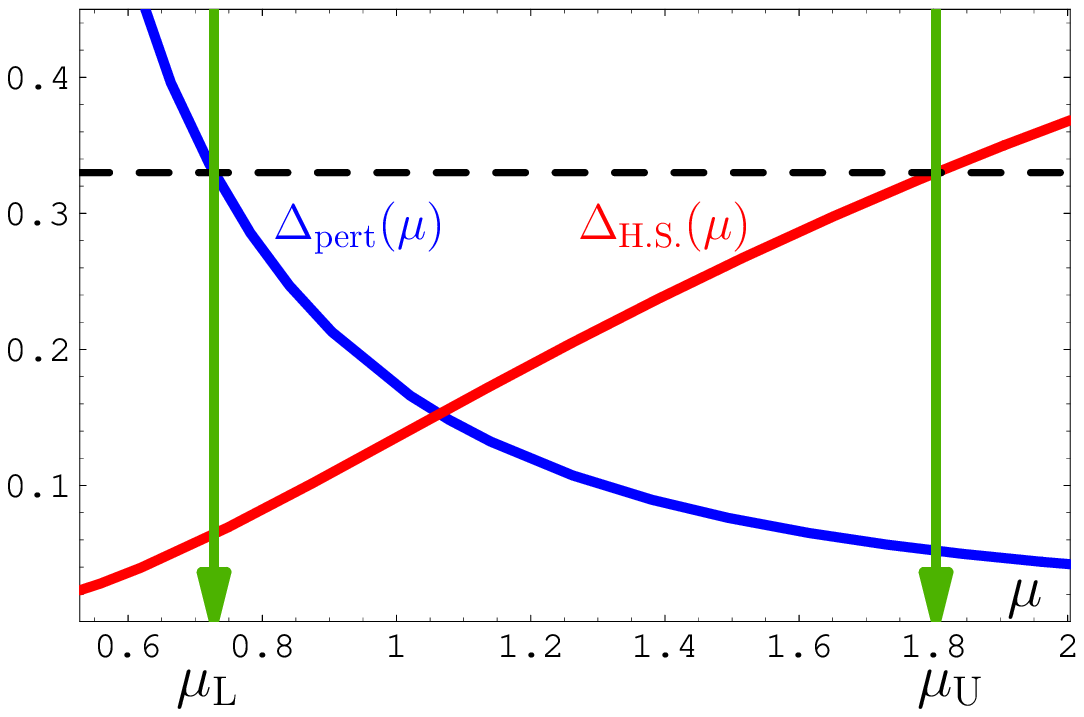}~%
             \includegraphics[width=0.49\textwidth]{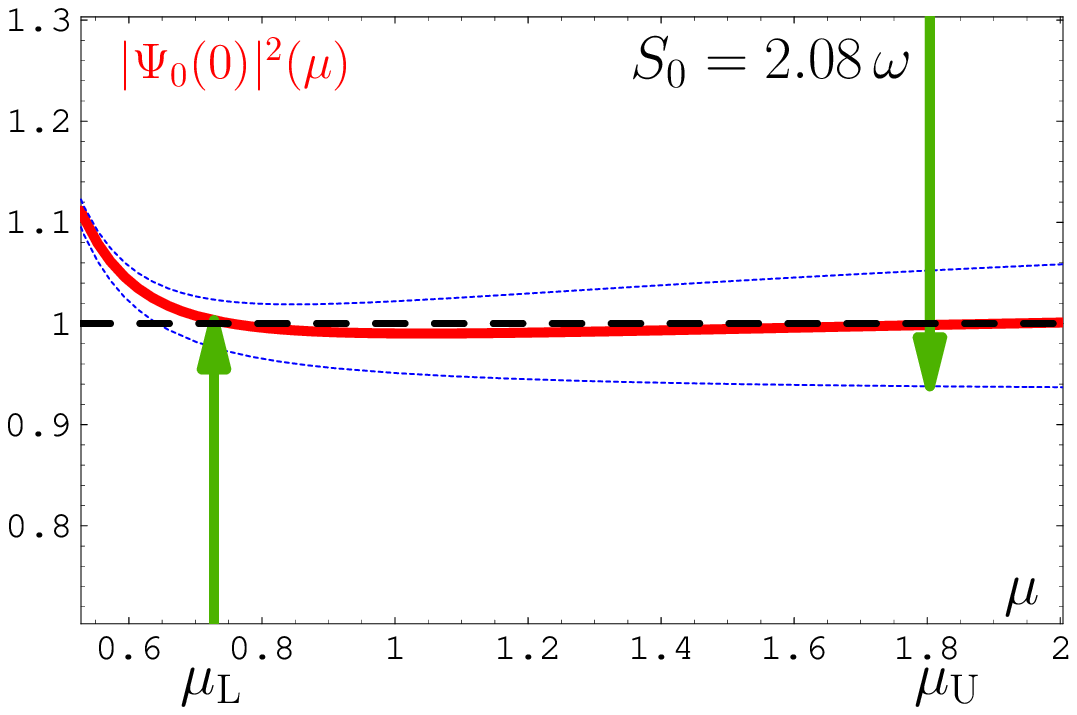}}
  \caption{\footnotesize \textbf{Left panel:} Determination of the fidelity window
  using criteria (\ref{eq:Delta.Pert.33}) and (\ref{eq:Delta.HS.33}).  
  \textbf{Right panel:} We show here the l.~h.~s. of Eq.\ (\ref{eq:SR1}) with 
  $E_0$ fixed at the exact value $E_0=\omega$
  as a function of $\mu$ (red solid line). 
  Black dashed line corresponds to the position of the exact value of $|\Psi_0(0)|^2$.
  Positions of $\mu_\text{L}$ and $\mu_\text{U}$ are shown 
  by green vertical arrows in both panels and $\mu$ is displayed in units 
  of $\omega$.\label{fig:SR1.E0.fixed}}
\end{figure}

Let us first consider \textit{SR setup with fixed $\bm{E_0}$}:
We fix the energy of the ground state to the exact value, 
$E_0=\omega$,
and obtain the following fidelity window,
${\mu_\text{L}=0.73\,\omega}$
and
${\mu_\text{U}=1.80\,\omega}$,
see Fig.\ \ref{fig:SR1.E0.fixed}, left panel.
The result $|\Psi_0(0)|^2=0.99$ is obtained with only 2 power corrections included 
and for $S_0=2.08\,\omega$ 
(the exact value is $|\Psi_0(0)|^2=1$),
see Fig.\ \ref{fig:SR1.E0.fixed}, right panel.
%%%%%%%%%%%%%%%%%%%%%%%%%%%%%%%%%%%%%%%%%%%%%%%%%%%%%%%%%%%%%%%%%%%%%%%%%%%%%%%%
%%%%%%%%%%%%%%%%%%%%%%%%%%%%%%%%%%%%%%%%%%%%%%%%%%%%%%%%%%%%%%%%%%%%%%%%%%%%%%%%

\begin{figure}[hb]
 \centerline{\includegraphics[width=0.49\textwidth]{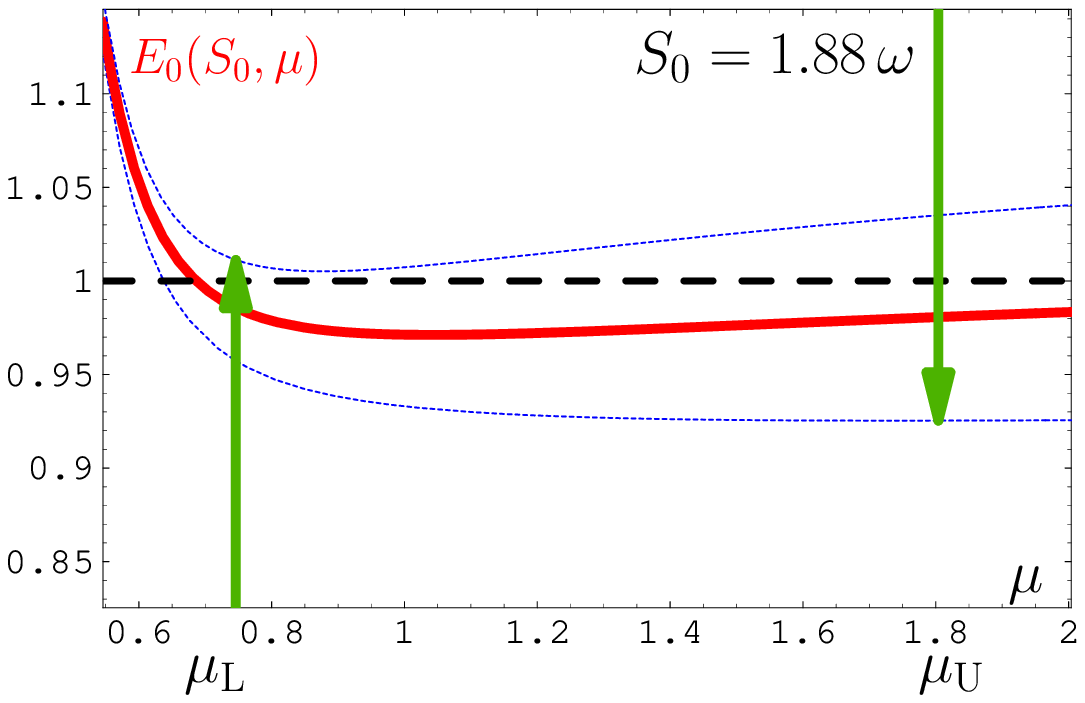}~%
             \includegraphics[width=0.49\textwidth]{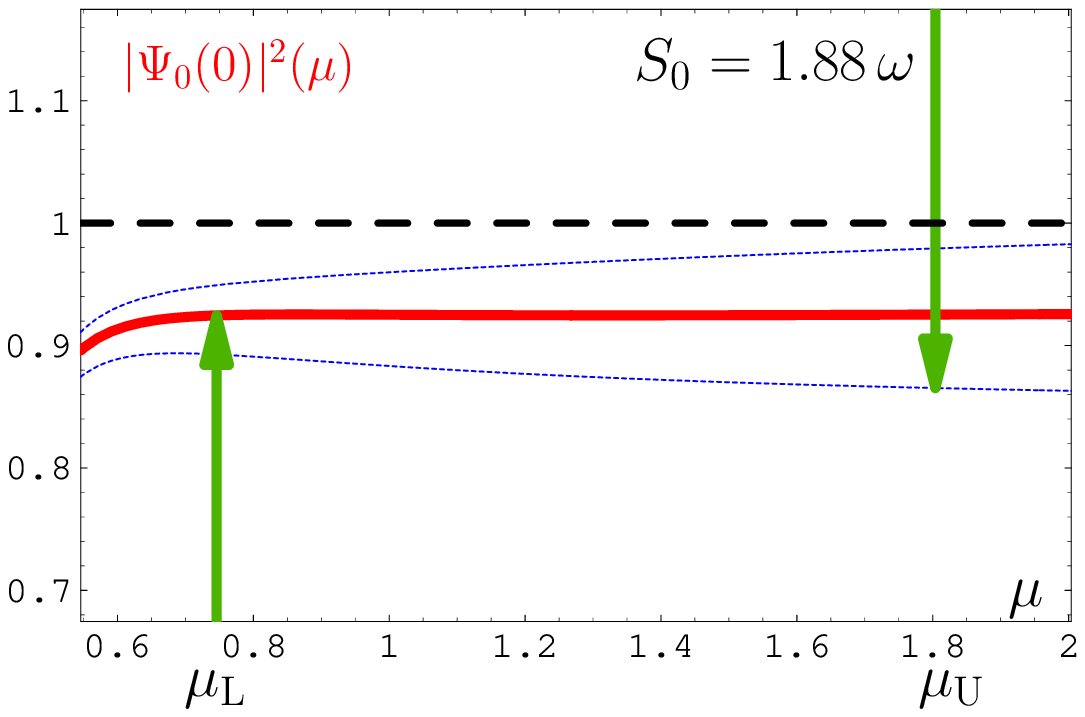}}
  \caption{\footnotesize \textbf{Left panel:} We show here the l.~h.~s. 
  of Eq.\ (\ref{eq:SR.E0}), $E_0(S_0,\mu)/\omega$, as a function of $\mu$ 
  in units of $\omega$ (red solid line). 
  Black dashed line corresponds to the position of exact value of $E_0$.
  \textbf{Right panel:} The l.~h.~s. of Eq.\ (\ref{eq:SR1}) is shown as a function 
  of $\mu$ in units of $\omega$ (red solid line). 
  Black dashed line corresponds to the position of exact value of $|\Psi_0(0)|^2$.
  \label{fig:SR12.Comp}}
\end{figure}

Now we consider \textit{SR in the complete setup},
that means that we determine the energy of the ground state
from the daughter SR (\ref{eq:SR.E0}).
We take into account 3 power corrections and obtain 
the fidelity window ${[0.74\,\omega;1.8\,\omega]}$
and ${E_0=0.98\,\omega}$ for ${S_0=1.88\,\omega}$,
see Fig.\ \ref{fig:SR12.Comp}, right panel.
%%%%%%%%%%%%%%%%%%%%%%%%%%%%%%%%%%%%%%%%%%%%%%%%%%%%%%%%%%%%%%%%%%%%%%%%%%%%%%%%
%%%%%%%%%%%%%%%%%%%%%%%%%%%%%%%%%%%%%%%%%%%%%%%%%%%%%%%%%%%%%%%%%%%%%%%%%%%%%%%%

Our conclusions about SRs in quantum mechanics
can be summarized as follows:
\begin{itemize}
\item {SRs} give ${E_0}$ and ${|\psi_0(0)|^2}$
      with accuracy \textit{not worse than 10\%}\,;
\item Main source of the error -- \textit{crude model} 
 for the spectral density of higher states: 
 even taking into account 10 power corrections
 we obtain  ${E_0=0.95\,\omega}$, ${S_0=1.79\,\omega}$, 
 and ${|\psi_0(0)|^2=0.89}$\,;
\item \textit{But}: If we know ${E_0=1}$ exactly 
 (say, from Particle Data Group), 
 then accuracy can be twice higher: 
 with taking into account 2 power corrections
 we obtain  ${S_0=2.08\,\omega}$ 
 and ${|\psi_0(0)|^2=0.99}$\,!
\item In {QCD} spectral density more close to perturbative!
\end{itemize}

%%%%%%%%%%%%%%%%%%%%%%%%%%%%%%%%%%%%%%%%%%%%%%%%%%%%%%%%%%%%%%%%%%%%%%%%%%%%%%%%%%%%%%
\section{QCD SRs: Way to study hadrons in nonperturbative QCD}
%%%%%%%%%%%%%%%%%%%%%%%%%%%%%%%%%%%%%%%%%%%%%%%%%%%%%%%%%%%%%%%%%%%%%%%%%%%%%%%%%%%%%%
In QCD we have a big problem: nobody knows how to analyze bound states.
The method of QCD SRs allows us to calculate properties of hadrons
(masses, decay constants, magnetic moments)
without considering hadronization or confinement issues.
It was invented in 1977 by Shifman, Vainshtein \& Zakharov (ITEP)~\cite{NOSVVZ77} 
in order to describe {$J/\psi$}-meson, the {$c\bar{c}$}-system,
discovered in 1974 in {$e^+e^-$}-annihilation at SPEAR (SLAC)
and, in parallel, in  {$p+Be$}-collisions at BNL.
In 1979 this method was applied to describe light hadrons in {massless QCD}~\cite{SVZ}.

\textit{Main idea}: 
to calculate correlators of hadron currents 
$\va{0|T\left[J_{1}(x)J_{2}(0)\right]|0}$
by two approaches and 
to obtain the SR as the result of the matching.
We start with the Fourier transformed 
correlator of two hadron currents
\begin{eqnarray}
 \label{eq:Corr.12}
  \Pi\left(Q^2\right)\text{Lorentz}_{12}
   \equiv i\!\int\!\e^{iqx}
    \left[\va{0|T\left[J_{1}(x)J_{2}(0)\right]|0}\right]\,dx\,,
\end{eqnarray}
where Lorentz$_{12}$ includes all Lorentz structures,
so that $\Pi\left(Q^2\right)$ is scalar, 
and use 
the dispersion integral representation
\begin{eqnarray}
 \Pi(Q^2) 
   = \int\limits_{0}^\infty \frac{\rho_{12}\left(s\right)\,ds}{s+Q^2}
   + \text{``subtractions''}\,.~~~
 \label{eq:Corr.DR}  
\end{eqnarray}
Then we apply Borel transform 
defined as
\begin{eqnarray}
 \label{eq:Borel}
  \Phi(M^2) = \hat{B}(Q^2\to M^2) \Pi(Q^2) 
    = \mathop{\text{lim}}\limits_{n\to\infty}
       \frac{(-Q^2)^n}{\Gamma(n)}\,
        \left[\frac{d^n}{dQ^{2n}}\Pi(Q^2)\right]_{Q^2=n M^2}\,.
\end{eqnarray}
Here we list the most important examples:
$$\begin{tabular}{c|cccc}
 \hline \hline 
  $\Pi(Q^2)\vphantom{^{\big|}_{\big|}}$ 
      & $C= \textsc{const}$
          & $\displaystyle C\,\log\left(Q^2/\mu^2\right)$
              & $\displaystyle 1/Q^{2n}$
                  & $\displaystyle 1/\left(s+Q^2\right)$
     \\ \hline
  $\Phi(M^2)\vphantom{^{\big|}_{\big|}}$
      & $0$
          & $-C$
              & $\displaystyle 1/\left(\Gamma(n)\,M^{2n}\right)$
                 & $\displaystyle e^{-s/M^2}/M^2$
     \\  \hline \hline 
  \end{tabular}
$$
For us the most important are the first and the last columns in this table:
we see that Borel transform kills ``subtractions'' 
and suppresses ``higher states'' 
(by the factor $\exp(-s/M^2)$ in the integrand)
in (\ref{eq:Corr.DR}):
\begin{eqnarray}
  B_{Q^2\to M^2}\left[\Pi(Q^2)\right]
   \equiv \Phi\left(M^2\right)
   = \int\limits_{0}^\infty  
      \rho_{12}\left(s\right)\,e^{-s/M^2}
       \frac{ds}{M^2}~~
\end{eqnarray}
In the \textit{1-st approach} we apply 
the Operator Product Expansion (OPE) 
with account for 
quark and gluon condensates in QCD vacuum
to obtain
\begin{eqnarray}
  \Phi\left(Q^2\right)
   = \Phi_\text{pert}\left(Q^2\right)
   + c_{GG}\,\frac{\va{(\alpha_s/\pi)GG}}{M^4}
   + c_{\bar{q}q}\,\frac{\alpha_s\va{\bar{q} q}^2}{M^6}\,.
\end{eqnarray}
Here $\va{(\alpha_s/\pi)\,G^a_{\mu\nu}G^{a\mu\nu}}=0.012~\text{GeV}^4$,
$\alpha_s\va{\bar{q} q}^2=0.0018~\text{GeV}^6$ has been determined in~\cite{SVZ}
and till nowadays are practically the same~\cite{IZ00}.\\
The \textit{2-nd approach} uses phenomenological saturation 
of spectral density by hadronic states, see (\ref{eq:rho.s.X}),
\begin{eqnarray}
 \rho_\text{had}\left(s\right)
  = f_{h}^2\delta\left(s-m_h^2\right)
  + \rho_\text{pert}\left(s\right)\theta\left(s-{{s_0}}\right)\,.
\end{eqnarray}
Our model is the ground state $h$~$+$~{continuum},
which starts from threshold $s=s_0$.
Then the SR is
\begin{eqnarray}
 \label{eq:SR}
  f_{h}^2\,e^{-m_h^2/M^2}\,
   =\,\int\limits_{0}^{s_0}\!
       \rho_\text{pert}(s)\,
        e^{-s/M^2}ds 
   + c_{GG}\frac{\va{\frac{\alpha_s}{\pi}GG}}{M^2}
   + c_{\bar{q}q}\frac{\alpha_s\va{\bar{q} q}^2}{M^4}\,.
\end{eqnarray}
\textit{Our aim}: 
to determine ${ f_{h}^2}$ and ${m_h}$ from this SR 
by calculating spectral density 
${\rho_\text{pert}(s)}$ 
and coefficients $ c_{GG}$ and $ c_{\bar{q}q}$
and by demanding stability of this SR 
in variable ${M^2\in[M^2_\textbf{L},M^2_\textbf{U}]}$
with appropriate value of $S_0$.    

%%%%%%%%%%%%%%%%%%%%%%%%%%%%%%%%%%%%%%%%%%%%%%%%%%%%%%%%%%%%%%%%%%%%%%%%%%%%%%%%%%%%%%
\section{Mesonic currents in QCD}
%%%%%%%%%%%%%%%%%%%%%%%%%%%%%%%%%%%%%%%%%%%%%%%%%%%%%%%%%%%%%%%%%%%%%%%%%%%%%%%%%%%%%%
Let us now write down the currents related to $\pi^\pm$-mesons in QCD:
\begin{eqnarray}
 \text{AV:} & & J_{\mu5}(x)
                = \bar{u}(x)\gamma_\mu\gamma_5 d(x)\,;
        \quad  J^{\dagger}_{\mu5}(x)
                = \bar{d}(x)\gamma_\mu\gamma_5 u(x)\,;\\
 \text{PS:} & & ~J_{5}(x)
                = i\,\bar{u}(x)\gamma_5 d(x)\,;\quad  
                ~~J^{\dagger}_{5}(x)
                = i\,\bar{d}(x)\gamma_5 u(x)\,.
\end{eqnarray}
Note that Dirac equation {$i\,\hat{D}\,q(x)\,=\,m_q\,q(x)$}
gives us the relation:
\begin{eqnarray}
 \label{eq:A.to.V}
  \partial^\mu J_{\mu5}(x) 
 = \left(m_u+m_d\right)\,J_{5}(x)\,.
\end{eqnarray}
The decay constant $f_\pi$ of the physical pion $\pi(P)$
is defined via
\begin{eqnarray}
 \label{eq:A.fpi}
 \va{0\big|J_{\mu5}(0)\big|\pi(P)}
  = i\,f_{\pi}\,P_\mu\,.
\end{eqnarray}
It was measured in the decay $\pi\to\mu\nu_\mu$ to be $f_\pi=132~\text{MeV}$.
Eq.\ (\ref{eq:A.to.V}) then gives 
\begin{eqnarray}
 \label{eq:PS.fpi}
  \va{0\big|J_{5}(0)\big|\pi(P)}
   = \frac{f_{\pi}\,m_\pi^2}{m_u+m_d}\,,
\end{eqnarray}
meaning that the pion reveals itself both in the axial and pseudoscalar
currents!

Currents related to $\rho^\pm$-mesons in QCD are
\begin{eqnarray}\label{eq:J.rho.pm}
 J_{\mu}(x)=\bar{u}(x)\gamma_\mu d(x)\,;
 \quad
 J^{\dagger}{_\mu}(x)=\bar{d}(x)\gamma_\mu u(x)\,.
\end{eqnarray}
The decay constant ${f_\rho}$ of physical ${\rho^\pm(P,\varepsilon)}$-meson
with polarization ${\varepsilon}$ and momentum ${P}$,
satisfying ${(P\,\varepsilon)=0}$ and ${(\varepsilon,\varepsilon)=-1}$,
is defined through
\begin{eqnarray}
 \label{eq:f.rho}
 \va{0\big|J_{\mu}(0)\big|\rho(P,\varepsilon)} 
  = f_{\rho}\,m_\rho\,\varepsilon_\mu\,.
\end{eqnarray}
Decay ${\rho^0\to e^+e^-}$ allows to measure $f_{\rho^0}=150~\text{MeV}$,
then from isospin symmetry we can deduce 
$f_{\rho^\pm}=\sqrt{2}\,f_{\rho^0}=210~\text{MeV}$.

\textit{Interesting question}: 
Why do we put $T$-product of currents in the definition (\ref{eq:Corr.12})
of correlators?
To answer it we consider 
the vector current correlator $\Pi_{\mu\nu}$ in more detail.
Lorentz invariance and vector current conservation dictate
\begin{eqnarray}
 \label{eq:Pi.Mu.Nu}
 \Pi_{\mu\nu}(q)
  =i\,\int\!\!d^4x\, e^{iqx} 
    \va{0\big|T\left[J^\mu(x)J_\nu(0)\right]\big|0}
  =\left[q_\mu\,q_\nu - g_{\mu\nu}\,q^2\right]\,\Pi(q)\,.
\end{eqnarray}
Inserting $\hat{1}$ in between currents,
we obtain
\begin{eqnarray}
 \Pi(q)&=&
  \frac{-i\,(2\pi)^3}{3q^2}\,\sum_{X(p)}
   \delta(\vec{p}-\vec{q})\,\theta(p_0)\, 
    \Big|\va{0\big|J_\mu(0)\big|X(p)}\Big|^2
 \nonumber\\
       &\times&
   \int_0^{\infty}\!\!dt\, 
    \left[e^{i(q_0-p_0)t} + e^{-i(q_0+p_0)t}\right]\,.
 \label{eq:Pi.V.V}
\end{eqnarray}
From Sokhotsky identity we have
\begin{eqnarray}
 \int_0^{\infty}\!\!dt\,e^{\pm i\alpha t}
 = \pi\,\delta(\alpha)
   \pm i\,{\cal P}\frac{1}{\alpha}\,,
\end{eqnarray}
that give us
\begin{eqnarray}
 \textbf{Im}\,\Pi(q^2)
  = -\pi\,\frac{(2\pi)^3}{3q^2}\,
      \sum_{X(p)}
       \delta(\vec{p}-\vec{q})\,
        \delta(p_0-|q_0|)\, 
         \Big|\va{0\big|J_\mu(0)\big|X(p)}\Big|^2
\end{eqnarray}
As a result we have  
\begin{eqnarray}
 \label{eq:Im.Pi.Rho.V}
  \frac{1}{\pi}\,\textbf{Im}\,\Pi(q^2) = \rho(q^2)\theta(|q_0|) 
  = \rho(q^2)
\end{eqnarray}
 with
\begin{eqnarray}
 \label{eq:Rho.V}
 \rho(q^2)\,\theta(q_0)
  = \frac{-(2\pi)^3}{3q^2}\,\sum_{X(p)}
     \delta^{(4)}(q-p)\,\theta(p_0)\Big|
      \va{0\big|J_\mu(0)\big|X(p)}\Big|^2\,.
\end{eqnarray}
Lorentz invariance dictates that $\rho(q^2)\geq 0$. Indeed,
\begin{eqnarray}
 \va{0\big|J^\mu(x)\big|X(p)} 
  = \left[A\,p_\mu + B\,\varepsilon_\mu\right]\,e^{-ipx}
\end{eqnarray}
with $p\cdot\varepsilon=0$, 
and therefore $\varepsilon\cdot\varepsilon=-1$.
From current conservation 
it follows $A=0$, i.~e. ($B=f_X\,m_X$, see (\ref{eq:f.rho}))
\begin{eqnarray}
 \va{0\big|J^\mu(x)\big|X(p)}
  \va{X(p)\big|J^{\dagger}_\mu(x)\big|0} 
 = \big|f_{X}\big|^2\,m_X^2 \varepsilon^2 
 = - \big|f_{X}\big|^2\,m_X^2 \leq 0\,.
\end{eqnarray}
This gives us
\begin{eqnarray}
 \label{eq:rho.s.X}
 \rho(s)=\sum_{X}\Big|f_X\Big|^2\,\delta(s-m_X^2) \geq 0\,.
\end{eqnarray}
Now we can say why we put ${T}$-product in correlators 
-- then spectral densities, 
defined only by real particles,
are Lorentz invariant 
and \textit{depend only on} ${\bm{q^2}}$\,!
Indeed, we see that the structure $e^{i(q_0-p_0)t} + e^{-i(q_0+p_0)t}$ 
in (\ref{eq:Pi.V.V}) appears 
due to the presence of the $T$-product 
in the definition of the correlator (\ref{eq:Pi.Mu.Nu}).
And just this structure generates as a result
$\theta(|q_0|)$ in (\ref{eq:Im.Pi.Rho.V}). 

\begin{figure}[h]
 \centerline{\includegraphics[width=0.46\textwidth]{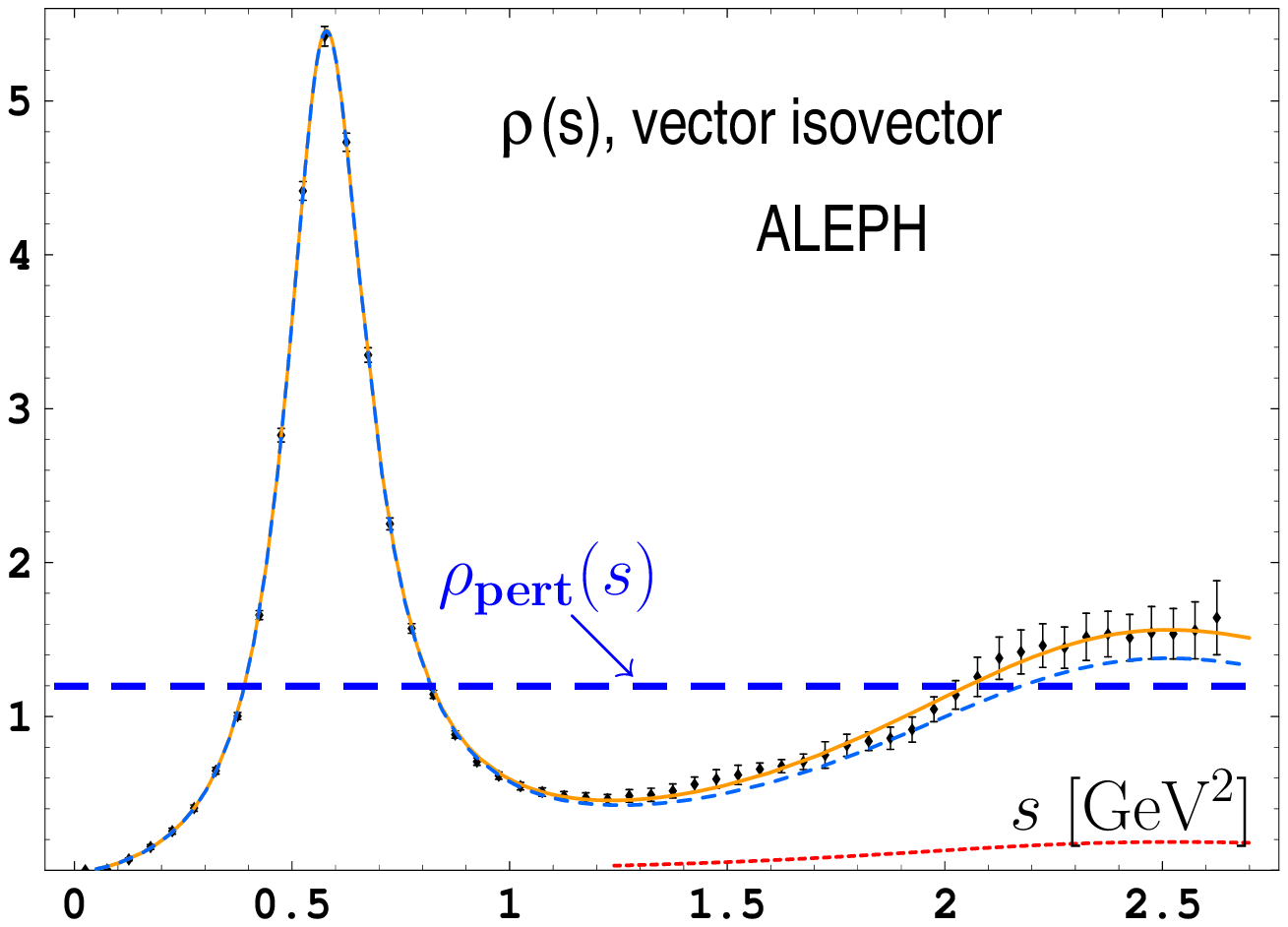}~%
             \includegraphics[width=0.53\textwidth]{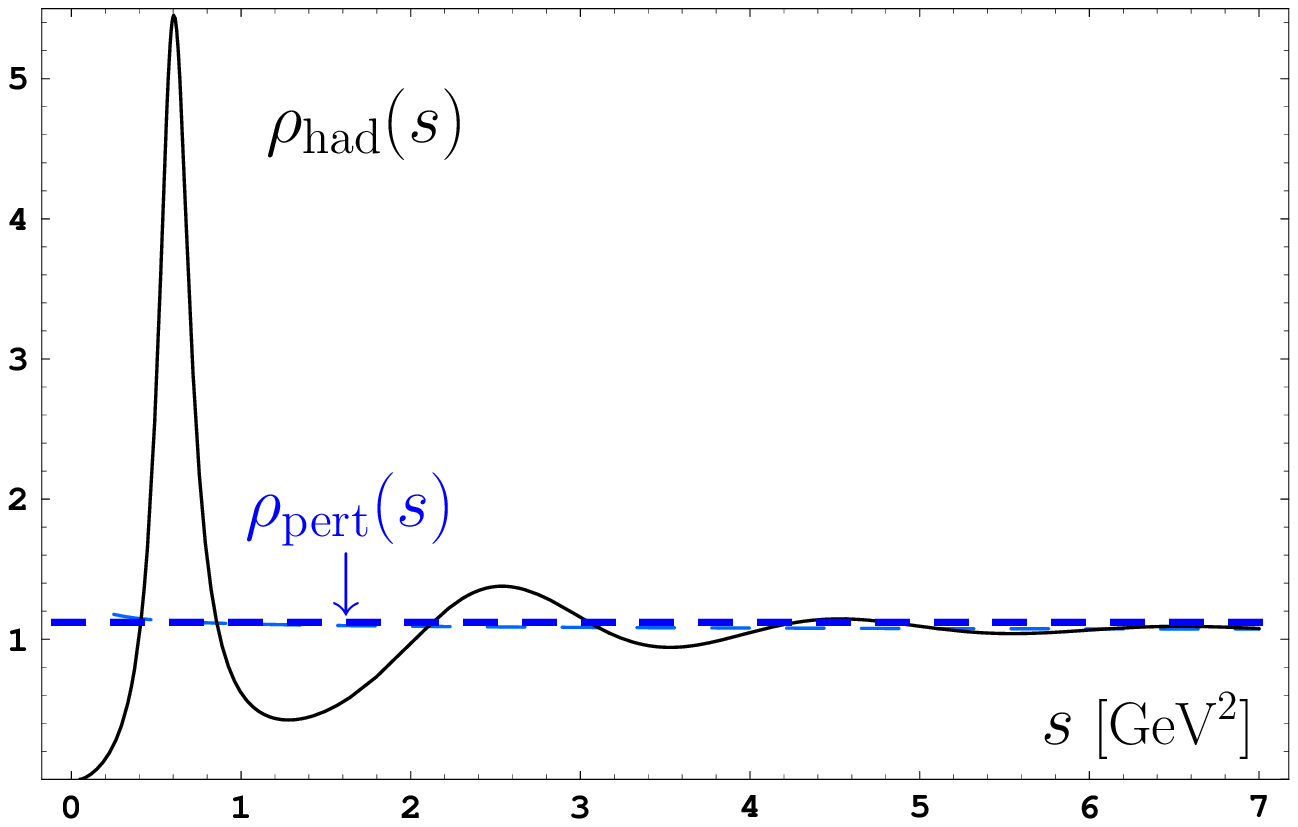}}
  \caption{\footnotesize We show here perturbative spectral density
   $4\pi^2\rho_\text{pert}(s)$ (blue dashed line) in comparison with spectral density
   $4\pi^2\rho_\text{had}(s)$ measured by the ALEPH Collaboration.\label{fig:ALEPH}}
\end{figure}

\textit{Relation to the cross section od $\bm{e^+e^-\!\to\!}$ hadrons}.
We have (\ref{eq:Rho.V}) for the spectral density $\rho(q^2)$. 
This function naturally appears in 1-photon QED description of
the process $e^+e^-\to$ hadrons 
if current $J_\mu$ is the electromagnetic current
due to quarks:\vspace*{-5mm}
$$\begin{minipage}[]{40mm}
 $$\includegraphics*[width=40mm]{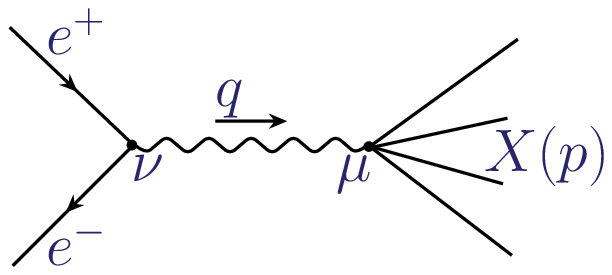}$$\end{minipage}~~~
  \begin{minipage}[]{60mm}\vspace*{6mm}
 {$\displaystyle\Leftrightarrow~~~\bar{u}(k)\gamma_\mu u(k')\frac{ie^2}{q^2}\va{X(p)|J_\mu(q)|0}$}
  \end{minipage}$$
\vspace*{-2mm}
Here $k$ and $k'$ are the momenta of ingoing $e^+$ and $e^-$.
Then we can deduce
\begin{eqnarray}
 \label{eq:sigma.e+e-}
  \sigma_{e^+e^-\to\text{hadrons}}(s)
  \,=\,\frac{16\,\pi^3\,\alpha^2}{s}\,\rho(s)
    =  \frac{4\,\pi\,\alpha^2}{3\,s}\,R(s)\,,
\end{eqnarray}
where we explicitly extracted as a factor 
the cross-section $\sigma_{e^+e^-\to \mu^+\mu^-}(s) = 4\,\pi\,\alpha^2/(3\,s)$
of the process $e^+e^-\to\mu^+\mu^-$.
Equivalently:
\begin{eqnarray}
 \label{eq:rho.R(s)}
  R(s)\,
   \equiv\,
   \frac{\displaystyle\sigma_{e^+e^-\to\text{hadrons}}(s)}{\displaystyle\sigma_{e^+e^-\to \mu^+\mu^-}(s)}\,;
 \quad
  \rho(s)\,
   =\, \frac{1}{12\,\pi^2}\,R(s)\,.
\end{eqnarray}
Now we can look to the quark-hadron duality,
that is duality of hadron spectral density $\rho_\text{had}(s)$,
which is measured in $\tau$-decay to $\nu_{\tau}+$hadrons,
and quark spectral density $\rho_\text{pert}(s)$
predicted by QCD 
(in this case $J_\mu$ is given by Eq.\ (\ref{eq:J.rho.pm})
 and in the leading order $R(s)=N_c$):
\begin{eqnarray}
 \label{eq:QHD}
  \int_{s_1}^{s_2}\!\! \rho_\text{pert}(s) ds
  = \int_{s_1}^{s_2}\!\! \rho_\text{had}(s) ds 
\end{eqnarray}
Looking in Fig.\ \ref{fig:ALEPH}
we can produce the following observations:
\begin{itemize}
  \item Real hadron spectral density is more smooth than in QHO case;
  \item Duality is working!
  \item Asymptotics starts at {${s\geq 3}~\text{GeV}{^2}$}
\end{itemize}
%%%%%%%%%%%%%%%%%%%%%%%%%%%%%%%%%%%%%%%%%%%%%%%%%%%%%%%%%%%%%%%%%%%%%%%%%%%%%%%%
%%%%%%%%%%%%%%%%%%%%%%%%%%%%%%%%%%%%%%%%%%%%%%%%%%%%%%%%%%%%%%%%%%%%%%%%%%%%%%%%

\section{Condensates in QCD}
In quantum mechanics we saw 
that in the presence of confinement potential 
$${M\left(\tau^{-1}\right)
        - M_0\left(\tau^{-1}\right)
      =  \frac{m}{2\pi}\,
          \left [- \frac{1}{6}\,\omega^2\,\tau 
          + \frac7{360}\,\omega^4\,\tau^3
          + \ldots
          \right]}\,.$$
This difference vanishes at short distances $\tau\ll1/\omega$ 
 and one can calculate  exact {$M(\mu)$} perturbatively, 
 expanding in powers of the oscillator potential.
In QCD \textit{confining potential} $V^\text{conf}(r)$ is not even known.
How to proceed further? The suggestion of QCD SR approach is:
 \begin{itemize}
  \item To construct perturbative expansion in terms of 
  quark and gluon propagators;
  \item To postulate that quark and gluon propagators
   \textit{are modified} by the long-range confinement potential;
  \item To suppose that this modification is soft: at $\tau\to0$
   the difference between \textit{exact} and \textit{perturbative} propagators 
   vanishes.
\end{itemize}
In realizing this program we write the exact propagator 
${\cal S}^\text{exact}(x,0)$
of a field $\psi$
as a vacuum average in the exact vacuum $|0\rangle$ 
\begin{eqnarray}
 \label{eq:S.Exact.Def}
  i\,{\cal S}^\text{exact}(0,x) = 
   \langle0|\,T(\bar{\psi}(x)\psi(0))|0\rangle\,.
\end{eqnarray}
Wick theorem allows us to write $T$-product 
as the sum
\begin{eqnarray}
 \label{eq:Wick}
  T(\bar{\psi}(x)\psi(0)) 
  =  \bar{\psi}(x)\BraSquare{-3}{12}{90}{13}\psi(0)
  + :\!\bar{\psi}(x)\,\psi(0)\!:
\end{eqnarray}
of the ``pairing'' and the ``normal'' product.
Then 
\begin{eqnarray}
 \label{eq:S.Exact.Start}
  {\cal S}^\text{exact}(0,x) 
  = S_0(x,0) 
  + \va{0|\!:\!\bar{\psi}(x)\,\psi(0)\!:\!|0}\,,
\end{eqnarray}
which can be considered 
as the starting point to calculate power corrections in QCD.
The examples in QCD are: 
$\va{\bar q q}\equiv\va{0|\!:\!\bar{q}(0)\,q(0)\!:\!|0}$ 
referred to as quark condensate;
$\va{\bar q D^2 q}$, characterizing average virtuality 
of the vacuum quarks;
gluon condensate 
$\va{GG}\equiv\va{0|\!:\!G_{\mu\nu}^a(0)\,G_{\mu\nu}^a(0)\!:\!|0}$,
\textit{etc.}
Here ${D_{\mu} \equiv \partial_{\mu} - igA_{\mu}}$ 
is the covariant derivative and 
${G_{\mu\nu} = (i/g)\bm{[}D_{\mu},D_{\nu}\bm{]}}$ 
is the gluonic field strength. 

\textit{Condensates and PCAC for pions in QCD}.
We derived the relations (\ref{eq:A.to.V}) and (\ref{eq:A.fpi})--(\ref{eq:PS.fpi}).
In order to see their consequences 
consider now correlator
\begin{eqnarray}
 \label{eq:Pi.A.PS}
  \Pi_{\mu55}(q)=i\int\!\!d^4x\, e^{iqx} 
   \va{0\big|T\left[J_{\mu5}(x)J^{\dagger}_5(0)\right]\big|0}
    \equiv i\,q_\mu\Pi_\text{AP}(q^2)
\end{eqnarray}
and its contraction with $q^\mu$
\begin{eqnarray}
 \label{eq:Pi.AP.q2.1}
 i\,q^2\,\Pi_\text{AP}(q^2)
 &=& -\int\!\!d^3\vec{x}\, e^{-i\vec{q}\vec{x}}
       \va{0\big|\bm{\left[}J_{05}(0,\vec{x});J^{\dagger}_5(0)\bm{\right]}\big|0}
 \nonumber\\
 &&- \left(m_u+m_d\right)
      \int\!\!d^4x\, e^{iqx}
       \va{0\big|T\left[J_{5}(x)J^{\dagger}_5(0)\right]\big|0}
 \nonumber\\
 &=& i\,\va{\bar{u}u+\bar{d}d}
  +  i\,\left(m_u+m_d\right)\,\Pi_{55}(q^2)\,.
\end{eqnarray}
Inserting pions in between currents of $\Pi_\text{AP}(q^2)$ in (\ref{eq:Pi.A.PS})
we have
\begin{eqnarray}
 \label{eq:Pi.AP.q2.2}
 \Pi_\text{AP}(q^2)
  \approx \frac{f_\pi\,m_\pi^2}{m_u+m_d}\,
                   \frac{f_\pi}{m_\pi^2-q^2}\Big|_{q^2\to\infty}
        = \frac{-f_\pi^2\,m_\pi^2}{m_u+m_d}\,
            \frac{1}{q^2}\left[1+O\left(\frac{m_\pi^2}{q^2}\right)\right]\,.~~~
\end{eqnarray}
Comparing asymptotics $O(1/q^2)$ of (\ref{eq:Pi.AP.q2.1}) and (\ref{eq:Pi.AP.q2.2}) gives us 
the famous \textit{PCAC relation}:
\begin{eqnarray}
 \label{eq:PCAC.pi}
 f_\pi^2\,m_\pi^2
  &=&-\,\va{\bar{u}u+\bar{d}d}
         \left(m_u+m_d\right)
     + O(m_q^2)\,.
\end{eqnarray}
In fact we should add other possible PS-meson states to obtain
\begin{eqnarray}
 \label{eq:PCAC}
 f_\pi^2\,m_\pi^2 + f_{\pi'}^2\,m_{\pi'}^2 + \ldots
  &=&-\,\va{\bar{u}u+\bar{d}d}
         \left(m_u+m_d\right)
     + O(m_q^2)\,.
\end{eqnarray}
In the chiral limit, $m_q\to0$, PCAC relation tells us:
\begin{itemize}
  \item $f_\pi\neq0$, then $m_\pi\to0$ $\Rightarrow$ pion is Goldstone boson;
  \item $m_{\pi'}\neq0$, then $f_{\pi'}\to0$ $\Rightarrow$ no decays $\pi'\to\mu\nu_\mu$\,!
  \item $m_{\pi}\approx f_{\pi}\approx130~\text{MeV}$ $\Rightarrow$ 
   $\va{\bar{q}q}\approx-(260~\text{MeV})^3$ at $m_u=m_d=4~\text{MeV}$. 
\end{itemize}

\section{QCD SRs for $\pi$-mesons}
 We study axial-axial correlator ${\Pi_{\mu5;\nu5}(q)}$
 $$ i\int\!\!d^4x\, e^{iqx} 
    \va{0\big|T\left[J_{\mu5}(x)J^{\dagger}_{\nu5}(0)\right]\big|0}
    \equiv g_{\mu\nu}\Pi_\text{1}(q^2)+q_\mu q_\nu \Pi_\text{2}(q^2)
 $$
Hadronic contribution to Borel transform of $\Pi_\text{2}(q^2)$:
 $$\Phi^\text{hadr}\left(M^2\right) 
           = B_{Q^2\to M^2}\left[\Pi^\text{hadr}_{2}(q^2)\right]
           = \frac{f_{\pi}^2}{M^2}
           + \frac{f_{A_1}^2}{M^2}\,e^{-m^2_{A_1}/M^2}
           + \ldots
 $$
The following diagrams contributes to the OPE of this correlator 
\newpage
$$\includegraphics[width=0.32\textwidth]{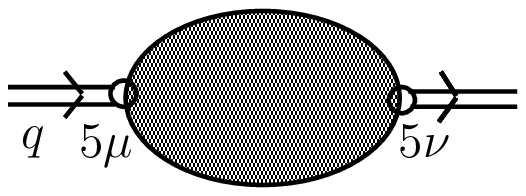}~%
  \begin{minipage}{0.32\textwidth}\vspace*{-11.9mm}%
   \includegraphics[width   =\textwidth]{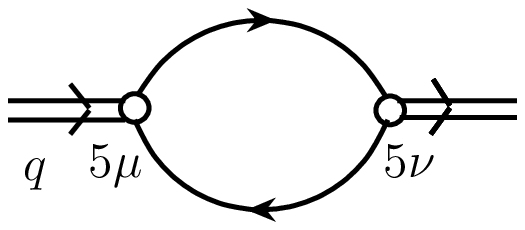}%
  \end{minipage}~%
  \begin{minipage}{0.32\textwidth}\vspace*{-14mm}%
   \includegraphics[width=\textwidth]{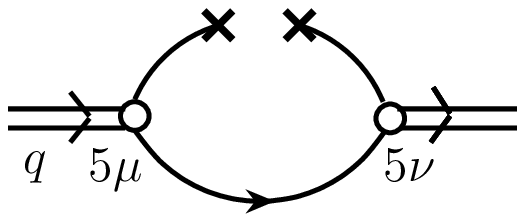}
  \end{minipage}\vspace*{-4mm}$$
$$\Phi_\text{OPE}\left(M^2\right)
  ~~~~~~=
  ~~~~~~\Phi_\text{pert}\left(M^2\right)
  ~~~~~~+
  ~~~~~~\Phi_\text{2V}\left(M^2\right)\vspace*{+2mm}
$$
$$\includegraphics[width=0.32\textwidth]{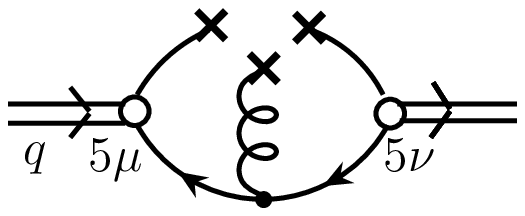}~%
  \begin{minipage}{\textwidth}\vspace*{-14mm}%
   \includegraphics[width=0.32\textwidth]{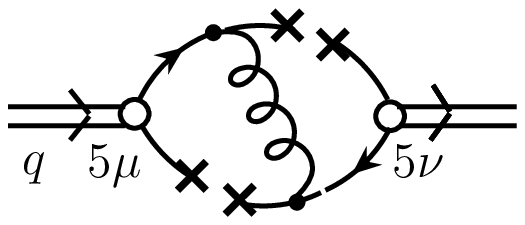}~%
  \end{minipage}~%
  \begin{minipage}{0.32\textwidth}\vspace*{-14mm}\hspace*{-84mm}%
  \includegraphics[width=\textwidth]{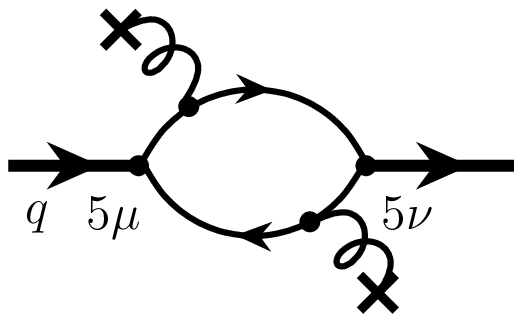}%
  \end{minipage}\vspace*{-2mm}$$
$$+~~~~\Phi_\text{3L}\left(M^2\right) 
  ~~~~~+
  ~~~~~\Phi_\text{4Q}\left(M^2\right)
  ~~~~~+
  ~~~~~\Phi_\text{GG}\left(M^2\right)~~~~~
$$
%%%%%%%%%%%%%%%%%%%%%%%%%%%%%%%%%%%%%%%%%%%%%%%%%%%%%%%%%%%%%%%%%%%%%%%%%%%%%%%%
\begin{figure}[b]
 \centerline{\includegraphics[width=0.49\textwidth]{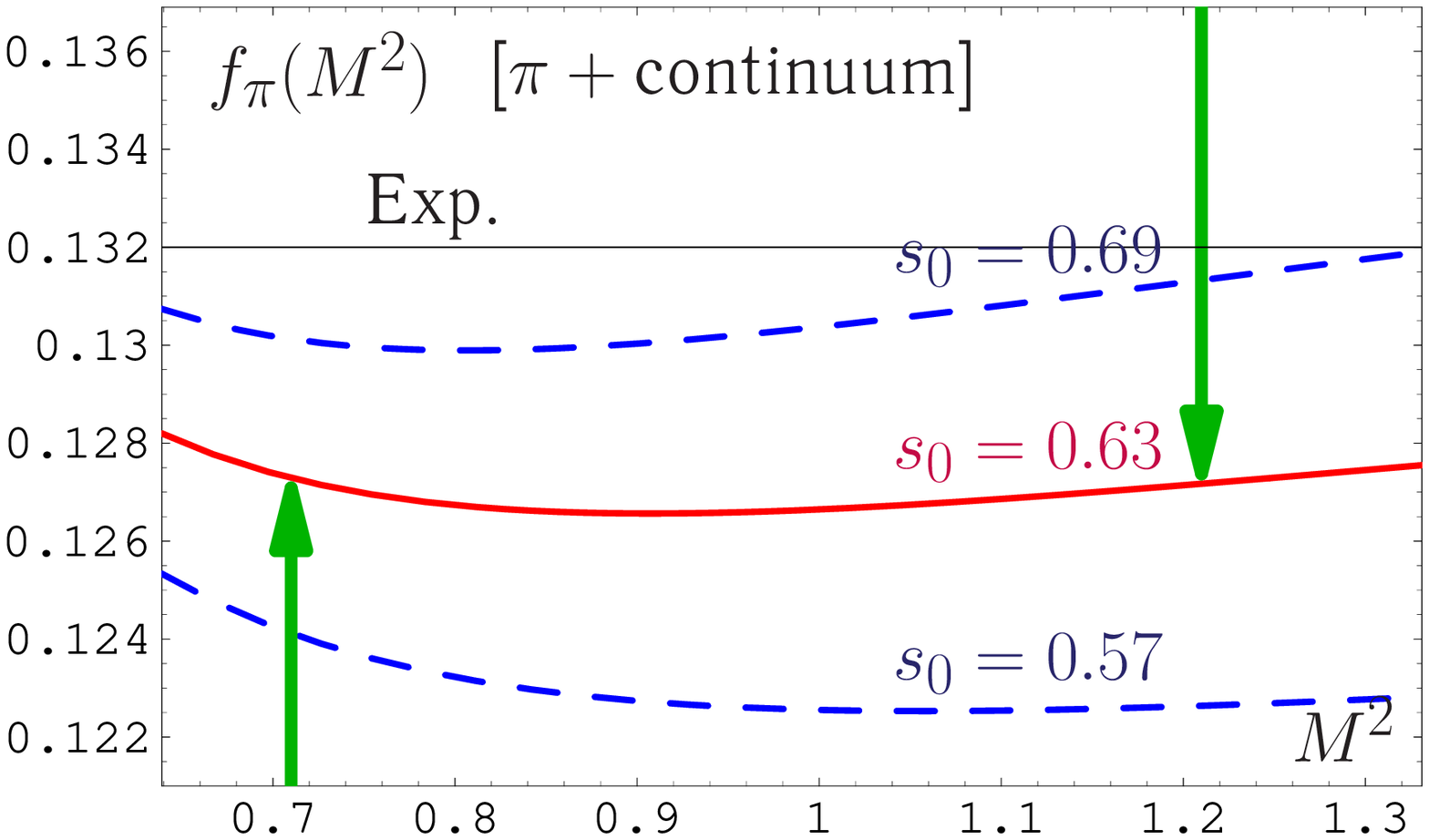}~%
             \includegraphics[width=0.49\textwidth]{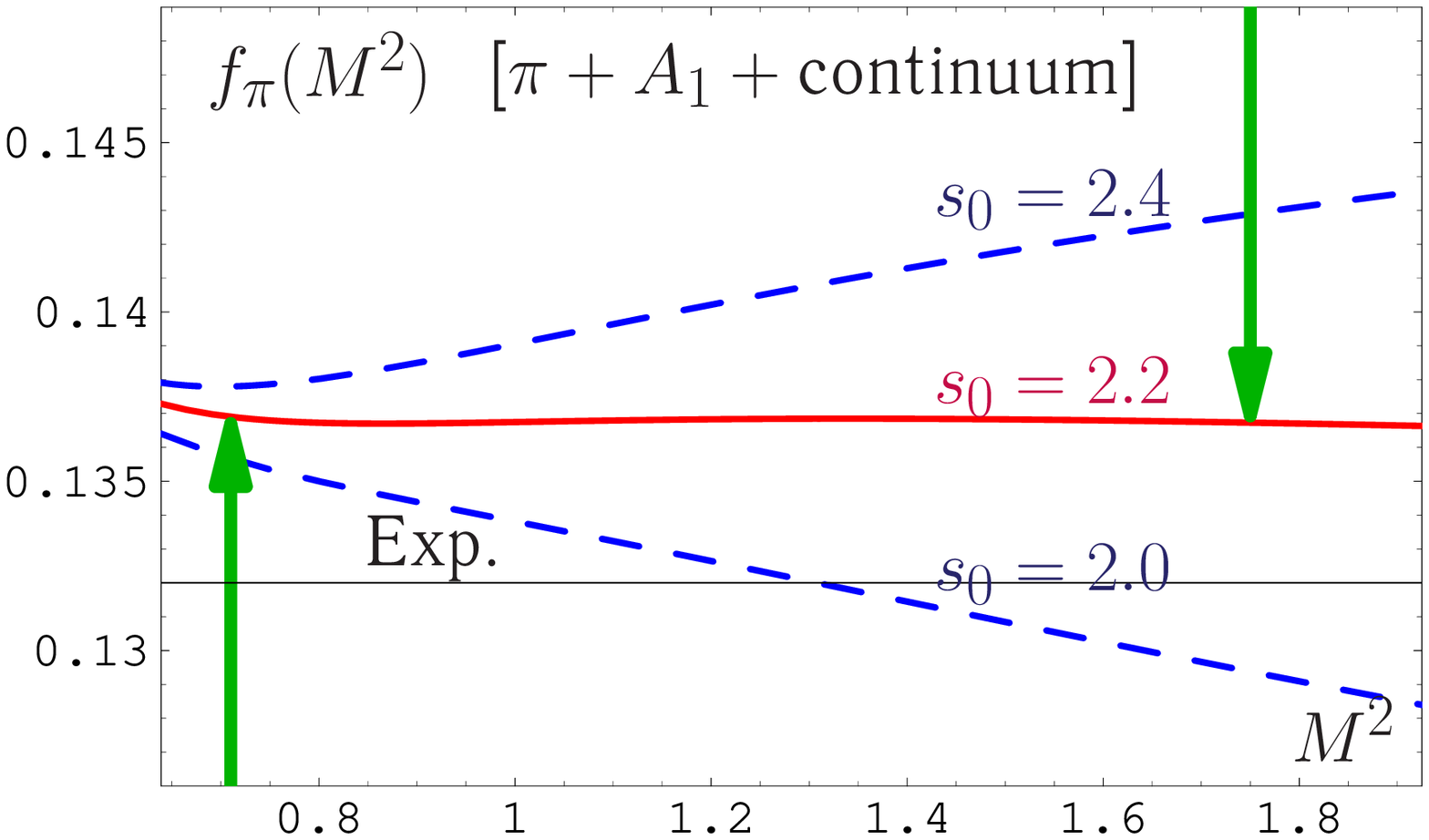}}
  \caption{\footnotesize \textbf{Left: panel} We show here the l.~h.~s. 
  of Eq.\ (\ref{eq:SR.fpi.pi}), $f_\pi^2(M^2)$, as a function of $M^2$ 
  (red solid line).  
  \textbf{Right panel:} The same as in the left panel, but for the spectral model 
  $\pi+A_1+\textsc{continuum}$.
  Blue dashed lines correspond to 10\% variations of the best tresholds.
  Positions of $M^2_\text{L}$ and $M^2_\text{U}$ are shown 
  by green vertical arrows in both panels.
  \vspace*{-3mm}\label{fig:fpi.SR}}
\end{figure}
%%%%%%%%%%%%%%%%%%%%%%%%%%%%%%%%%%%%%%%%%%%%%%%%%%%%%%%%%%%%%%%%%%%%%%%%%%%%%%%%
\noindent 
with 
$\Phi_{\{\text{2V},\text{3L},\text{4Q}\}}(M^2)$
$=\{16, 16, 144\}\,(\pi\alpha_s\va{\bar{q}q}^2)/(81\,M^6)$.
We see that in quark condensate contribution 
the most important one is $\Phi_{\text{4Q}}$.
As a result we have the following SR for the pion decay constant
\begin{eqnarray}
 \label{eq:SR.fpi.pi}
  f_{\pi}^2
   = \frac{M^2}{4\pi^2}
      \left(1-e^{-s_0/M^2}\right)
       \left[1+\frac{\alpha_s}{\pi}\right]
   + \frac{1}{12\,\pi}\ 
      \frac{\va{\alpha_s\,GG}}{M^2}
   + \frac{176}{81}\ 
      \frac{\pi\alpha_s\va{\bar{q}q}^2}{M^4}\,.~~~
\end{eqnarray}
Numerically, as can be seen from Fig.\ \ref{fig:fpi.SR}, 
we obtain $f_\pi=0.128\pm 0.13$~GeV from this SR,
whereas in the spectral model with $A_1$-meson
we obtain slightly higher value $f_\pi=0.137\pm 0.13$~GeV, 
to be compared with {$f_\pi^\text{exp}=0.132$~GeV}.

\section{Generalized QCD SRs for mesonic distribution amplitudes}
The \textit{pion distribution amplitude} (DA) parameterizes
the matrix element of the nonlocal axial current on the light cone~\cite{Rad77}
\begin{eqnarray}
 \label{eq:pion.DA.ME}
 \va{0\!\mid\!\bar d(z)\gamma_{\mu}\gamma_5 
  E(z,0)u(0)\!\mid\!\pi(P)}\Big|_{z^2=0}
  = i f_{\pi}P_{\mu}\!
       \int\limits_{0}^{1}\! dx\ e^{ix(zP)}
        \varphi_{\pi}^\text{Tw-2}(x,\mu^2)\,.~~
\end{eqnarray}
Here the gauge-invariance is guarantied by the Fock--Schwinger string
$$E(z,0)={\cal P}\exp\left[i g\int_0^z A_\mu(\tau) d\tau^\mu\right]$$ 
in between separated quark fields.
The physical meaning of this DA --- the amplitude of the transition 
$\pi(P)\rightarrow u(Px) + \bar{d}(P(1-x))$.
It is convenient to represent the pion DA:
\begin{eqnarray}
 \label{eq:pi.DA.rep}
  \varphi_\pi(x;\mu^2) 
  = \varphi^\text{As}(x)\,
     \Bigl[1 + \sum\limits_{n\geq1}a_{2n}(\mu^2)\,C^{3/2}_{2n}(2x-1)
     \Bigr]\,,
\end{eqnarray}
where $C^{3/2}_n(2x-1)$ are the {Gegenbauer} polynomials
(1-loop eigenfunctions of ER-BL kernel)
and $\varphi^\text{As}(x)=6\,x\,(1-x)$.
This representations means 
that all scale dependence  in $\varphi_\pi(x;\mu^2)$
is transformed to the scale dependence of the set 
$\left\{a_2(\mu^2), a_4(\mu^2), \ldots\right\}$.
ER-BL solution at the 2-loop level
is also possible 
with 
using the same representation 
(\ref{eq:pi.DA.rep})~\cite{MR86ev,KMR86,Mul94,BS05}.

\begin{figure}[t]
 \begin{tabular}{cc}
  \begin{minipage}{0.49\textwidth}
   $$\includegraphics[width=\textwidth]{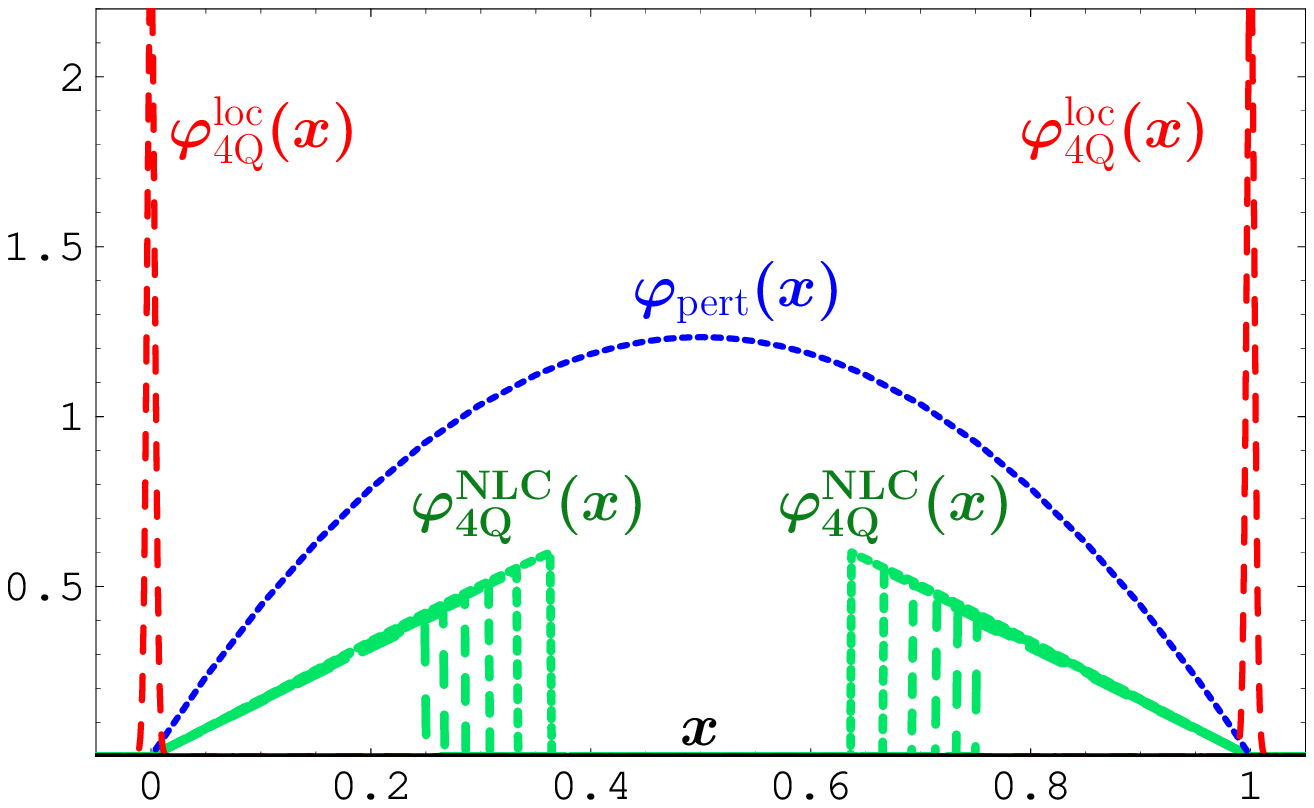}$$
  \end{minipage}&
  \begin{minipage}{0.49\textwidth}
   $$\includegraphics[width=\textwidth]{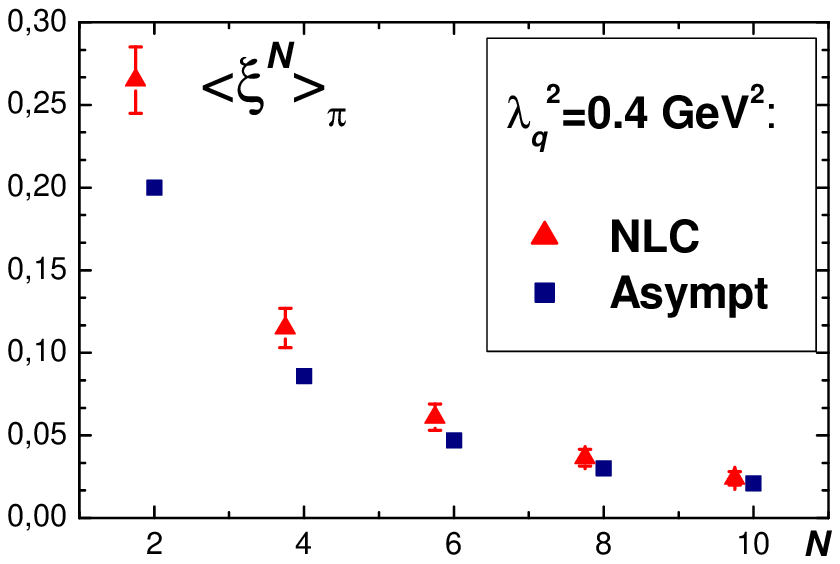}$$
   \end{minipage} 
 \end{tabular}
 \caption{\footnotesize \textbf{Left panel}: We show here contributions 
  to the R.~H.~S. of Eq.\ (\ref{eq:NLC.SR.pion.DA})
  due to perturbative loop (blue dotted line)
  and due to 4Q-condensate: $\varphi^\text{loc}_\text{4Q}(x)$
  -- in standard QCD SRs, and  $\varphi^\text{NLC}_\text{4Q}(x,M^2)$ 
  (with $M^2=0.55-0.80$~GeV$^2$) -- in NLC QCD SRs.
  \textbf{Right panel}: Moments $\va{\xi^{N}}_\pi$  
  obtained using the NLC SR (\ref{eq:NLC.SR.pion.DA}), 
  are shown by triangles with error-bars. We show also for comparison
  the asymptotic DA moments (squares).
  \label{fig:piDA.SR.4Q}\vspace*{-4mm}}
\end{figure}

In order to construct reliable SRs for the pion DA
moments 
one needs, as has been shown in~\cite{MR86,BM98},
to take into account 
the \textit{nonlocality of QCD vacuum condensates}.
For an illustration of the nonlocal condensate (NLC) model
we use here the minimal Gaussian model
\begin{eqnarray}
 \label{eq:Min.Gauss.Mod}
  \va{\bar{q}(0)q(z)} 
   = \va{\bar{q}\,q}\,
      e^{-|z^2|\lambda_q^2/8}\,.
\end{eqnarray} 
The single scale parameter $\lambda_q^2 = \va{k^2}$
characterizes 
the average momentum of quarks in the QCD vacuum
and has been estimated in QCD SR approach 
and also on the lattice\cite{BI82lam,OPiv88,Piv91,DDM99,BM02}:
\begin{eqnarray}
 \label{eq:lambda.q.SR}
  \lambda_q^2
   = 0.35-0.55~\text{GeV}^2\,.
\end{eqnarray}  
That means that $\lambda_q^2$ is of an order of the typical hadronic scale
~$m_{\rho}^2 \approx 0.6$ GeV$^2$. 
We write down as an example 
the NLC QCD SR for the pion DA itself.
To produce it one starts 
with a correlator of currents
$J_{\mu5}(x)$ and 
$J^{\dagger}_{\nu5;N}(0)=\bar{d}(0)\,\hat{n}\,\gamma_5\left(n\nabla\right)^N\!u(0)$
with light-like vector $n$, $n^2=0$,
obtains SRs for the moments $\va{x^N}_{\pi}$
and then realizes inverse Mellin transform 
from the moments $\va{x^N}_{\pi}$ to the DA $\varphi_\pi(x)$:
\begin{eqnarray}
 \label{eq:NLC.SR.pion.DA}
   f_{\pi}^2\,\varphi_\pi(x) 
   &=& \int_{0}^{s_{\pi0}}\rho^\text{pert}(x;s)\,
         e^{-s/M^2}ds 
     + \frac{\langle(\alpha_s/\pi)GG\rangle}{24 M^2}\,
        \varphi_{GG}(x;\Delta)\nonumber\\
   & &~~~~~~~~~~~~~~~~~~~
     +~\frac{8\pi\alpha_s\va{\bar{q}q}^2}{81M^4}
        \sum_{i=2V,3L,4Q}\varphi_i(x;\Delta)\,.~~~
\end{eqnarray}
The local limit $\lambda_q^2/M^2\equiv\Delta\to0$ of this SR 
is specified by the appearance of $\delta$-functions 
concentrated at the end-points $x=0$ and $x=1$,
for example,
$\varphi_{GG}(x;\Delta)=[\delta(x)+\delta(1-x)]$
and 
$\varphi_{4Q}(x;\Delta)=9[\delta(x)+\delta(1-x)]$. 
The minimal Gaussian model (\ref{eq:Min.Gauss.Mod})
generates the contribution $\varphi_\text{4Q}(x;\Delta)$
shown in the left panel of Fig.\ \ref{fig:piDA.SR.4Q}
in comparison with perturbative one
for the standard (local) and the NLC types of the SR.
We see that in the local version
due to completely different behaviour of perturbative and condensate
terms it is difficult to obtain some kind of consistency.
Alternatively, the NLC contribution is much more similar to the perturbative
one --- and for this reason in the NLC SR we have a very good stability!
After processing the SR (\ref{eq:NLC.SR.pion.DA}) for the moments 
(at $\mu^2\simeq1.35$ GeV$^2$)
\begin{eqnarray}
 \label{eq:piDA.moments}
  \va{\xi^N}_\pi 
   = \int_0^1\varphi_\pi(x)\,
      \left(2x-1\right)^N dx\,,
\end{eqnarray}
%%%%%%%%%%%%%%%%%%%%%%%%%%%%%%%%%%%%%%%%%%%%%%%%%%%%%%%%%%%%%%%%%%%%%%%%%%%%%%%%
%%%%%%%%%%%%%%%%%%%%%%%%%%%%%%%%%%%%%%%%%%%%%%%%%%%%%%%%%%%%%%%%%%%%%%%%%%%%%%%%
\noindent
we restore the pion DA $\varphi_\pi(x)$ 
by demanding that it should reproduce these 5 moments
and applying the minimally possible number of the Gegenbauer harmonics
in representation (\ref{eq:pi.DA.rep}).
It appears the NLC SRs for the pion DA
produce a \textit{bunch of self-consistent 2-parameter models} 
at $\mu^2\simeq 1.35$ GeV$^2$:
\begin{eqnarray}
 \label{eq:pi.DA.2Geg}
  \varphi^\text{NLC}_\pi(x;\mu^2) 
  = \varphi^\text{As}(x)\,
     \Bigl[1 + \sum\limits_{n=1,2}a_{2n}(\mu^2)\,C^{3/2}_{2n}(2x-1)
     \Bigr]\,.~
\end{eqnarray}
%%%%%%%%%%%%%%%%%%%%%%%%%%%%%%%%%%%%%%%%%%%%%%%%%%%%%%%%%%%%%%%%%%%%%%%%%%%%%%%%
\begin{figure}[t]\hspace*{-3mm}
 \begin{tabular}{cc}
  \begin{minipage}{0.48\textwidth}
   $$\includegraphics[width=\textwidth]{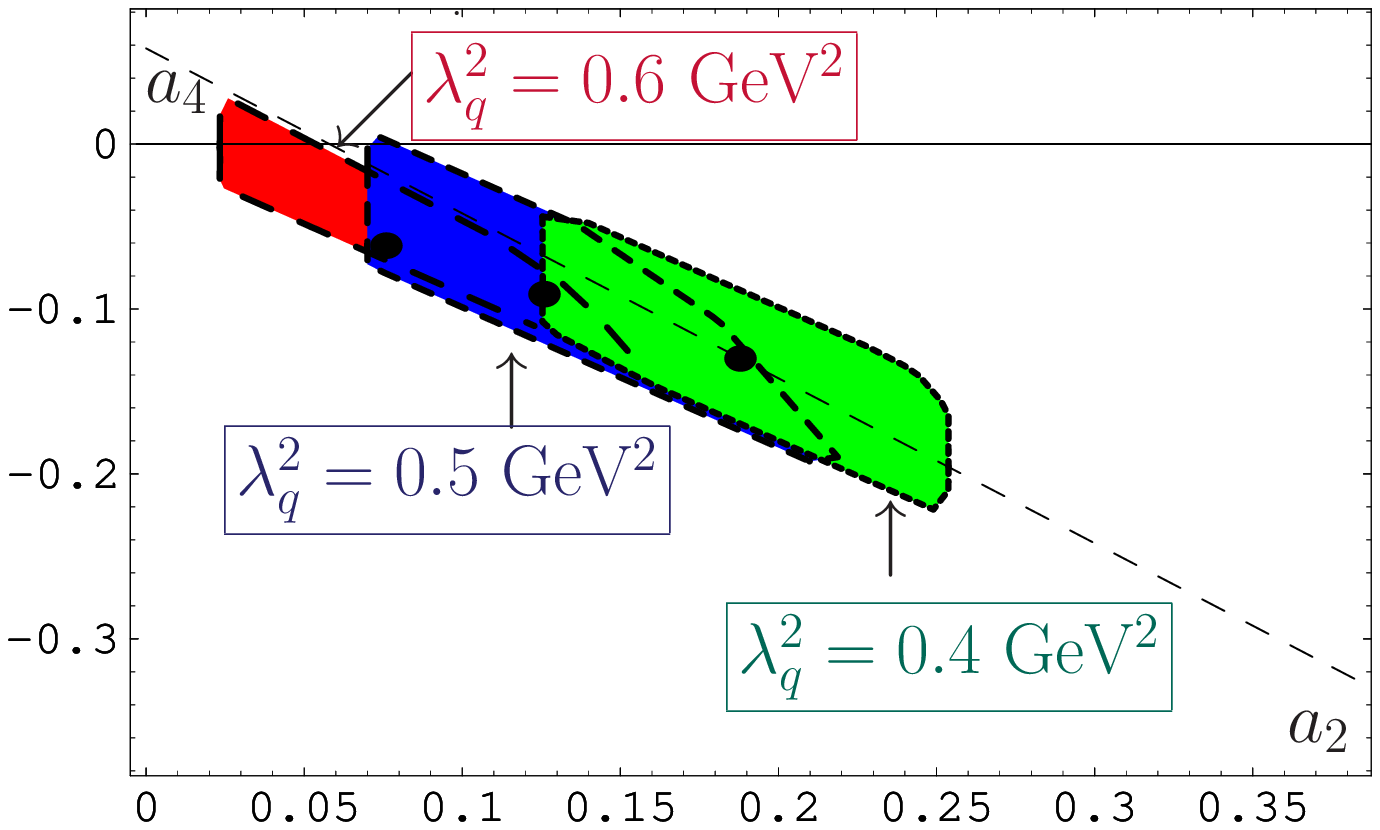}$$
  \end{minipage}&
  \begin{minipage}{0.48\textwidth}
   $$\includegraphics[width=\textwidth]{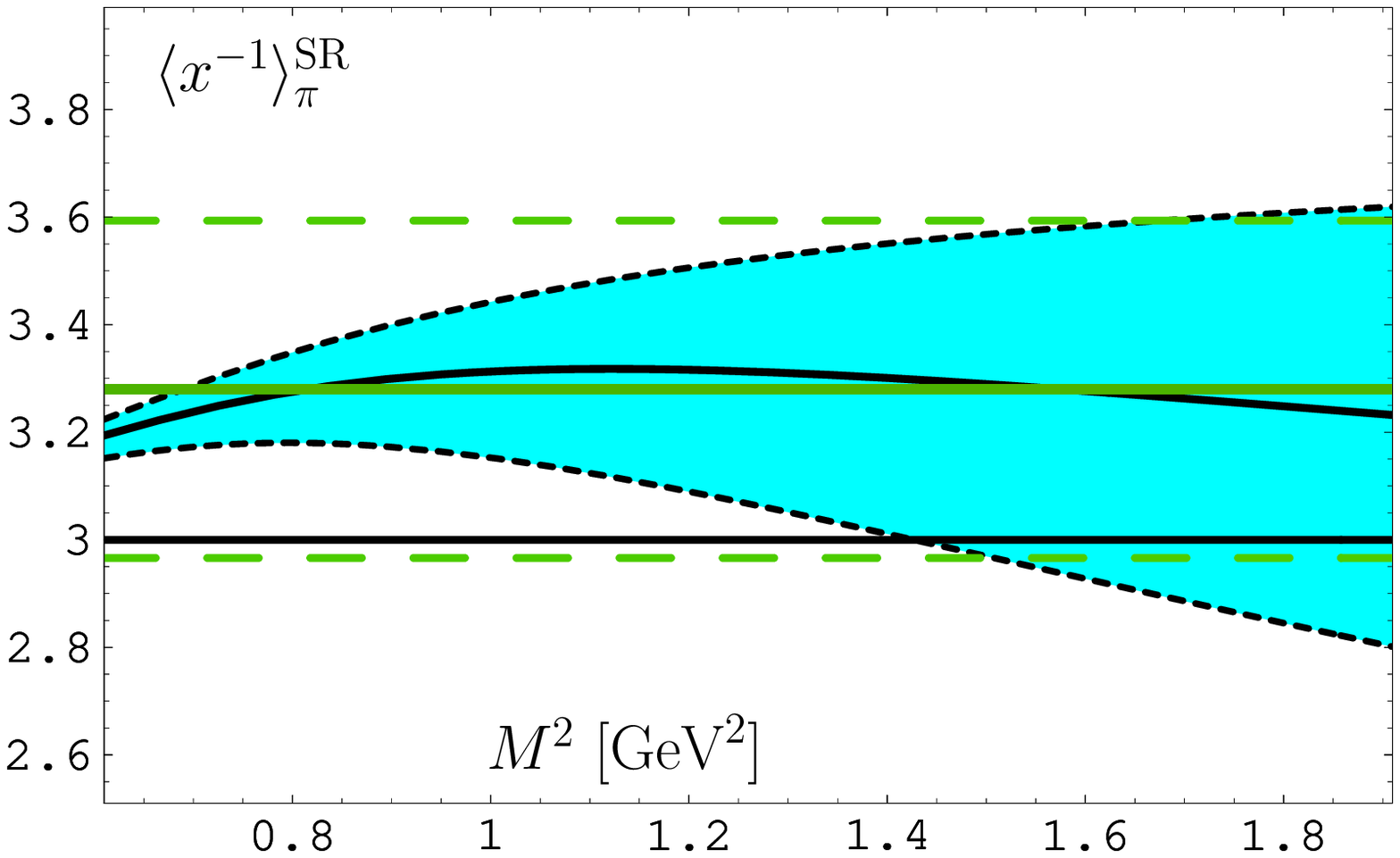}$$
   \end{minipage} 
 \end{tabular}
 \caption{\footnotesize \textbf{Left panel}: The allowed values 
 of parameters $a_2$ and $a_4$ of the bunches (\ref{eq:pi.DA.2Geg})
 at $\mu^2=1.35$~GeV$^2$
 for three values of the nonlocality parameter $\lambda_q^2$:
 $0.4$~GeV$^2$ (green region), $0.5$~GeV$^2$ (blue region), 
 and $0.6$~GeV$^2$ (red region).
  \textbf{Right panel}: The inverse moment $\va{x^{-1}}_\pi$, 
  obtained using the NLC SR (\ref{eq:NLC.SR.pion.DA}), 
  is shown by the green solid line (central value) with error-bars,
  shown as green dashed lines.
  \vspace*{-3mm}\label{fig:456}}
\end{figure}
%%%%%%%%%%%%%%%%%%%%%%%%%%%%%%%%%%%%%%%%%%%%%%%%%%%%%%%%%%%%%%%%%%%%%%%%%%%%%%%%
The central point corresponds to $a_2^\text{BMS}=+ 0.188$, $a_4^\text{BMS}=-0.130$
in the case $\lambda^2_q=0.4$ GeV$^2$,
whereas other allowed values of parameters $a_2$ and $a_4$
are shown in the left panel of Fig.\ \ref{fig:456} 
as the green slanted rectangle~\cite{BMS01}.
We verify that this solution is self-consistent
by estimating the inverse moment of the pion DA,
$\va{x^{-1}}_\pi$, 
in two ways.
The first is based on (\ref{eq:pi.DA.2Geg}) and gives us
\begin{eqnarray}
 \label{eq:pion.DA.Inv.Mom}
  \va{x^{-1}}^\text{bunch}_{\pi} = 3.17\pm0.20\,.
\end{eqnarray}
The second way uses the special SR for this moment,
obtained through the basic SR (\ref{eq:NLC.SR.pion.DA}).
It is worth to emphasize here that the moment 
$\va{x^{-1}}^\text{SR}_{\pi}$ could be determined 
only in NLC SRs 
because the end-point singularities absent.
This SR produces the estimate, 
see Fig.\ \ref{fig:456}, right panel:
\begin{eqnarray}
 \label{eq:Inv.Mom.SR}
 \va{x^{-1}}^{\text{SR}}=3.30\pm0.30
\end{eqnarray}
at $\mu^2\simeq 1.35$ GeV$^2$.
We see that both estimates are in a good agreement.

\begin{figure}[h]\hspace*{-3mm}
 \begin{tabular}{cc}
  \begin{minipage}{0.49\textwidth}
   $$\includegraphics[width=\textwidth]{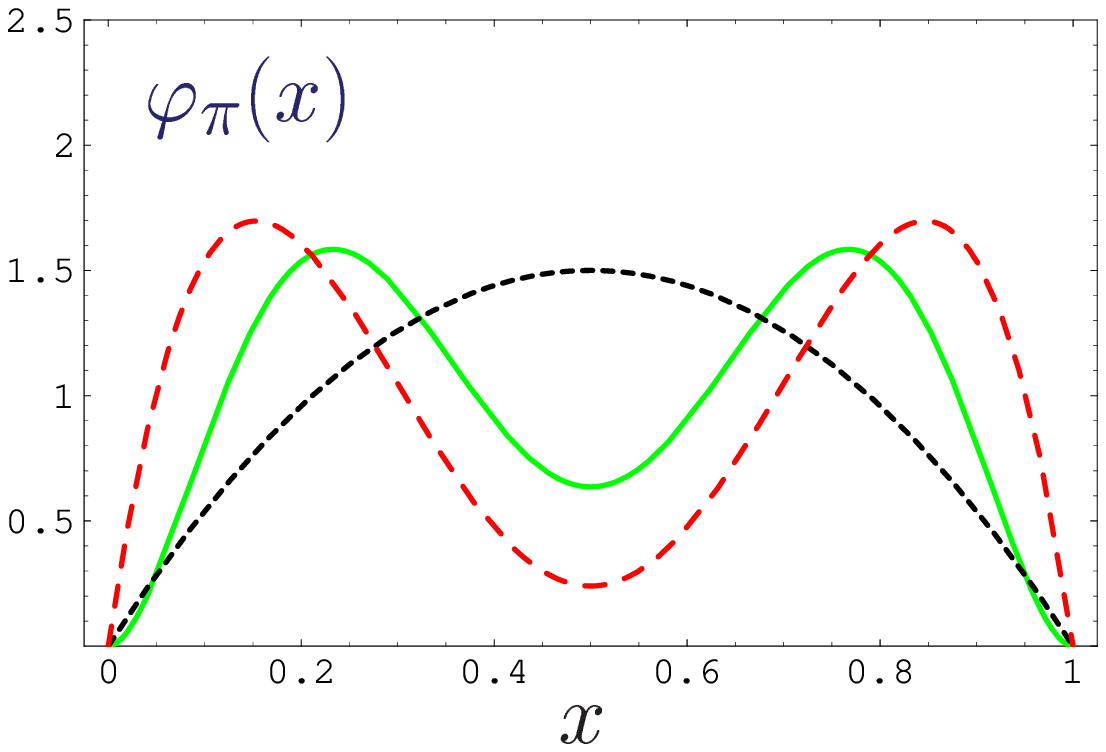}$$
  \end{minipage}&
  \begin{minipage}{0.49\textwidth}
   $$\includegraphics[width=\textwidth]{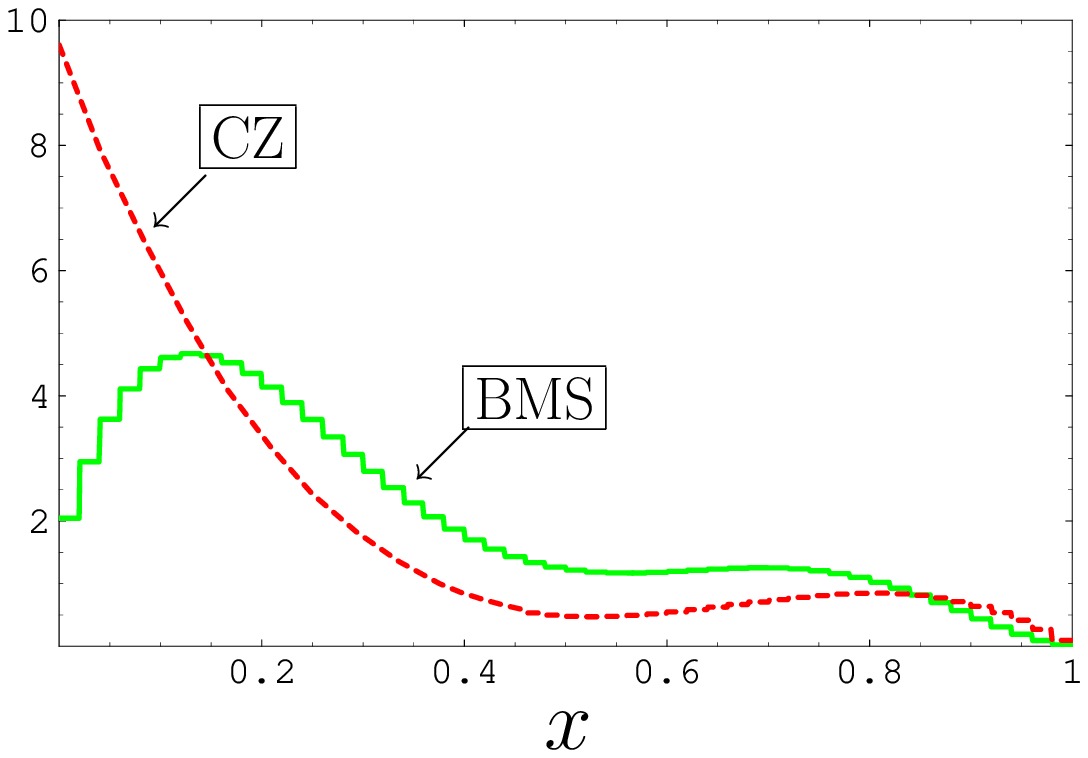}$$
   \end{minipage} 
 \end{tabular}
 \caption{\footnotesize \textbf{Left panel}: We show here 
  the comparison of curves for three DAs --- BMS (green solid line),
  CZ (red dashed line), and the asymptotic DA (black dotted line).
  \textbf{Right panel}: Histograms for contributions of different bins 
  to inverse moment $\va{x^{-1}}_{\pi}$ are shown for CZ and BMS DAs.
  \label{fig:CZ.As.BMS}}
\end{figure}

Comparing the obtained pion DA with 
the Chernyak\&Zhitni\-tsky (CZ) one~\cite{CZ82}, 
see Fig.\ \ref{fig:CZ.As.BMS},
reveals that although both DAs are two-humped
they are quite different:
our DA is strongly end-point suppressed.
This can also be verified 
in the right panel of the figure,
where contributions of different bins 
to inverse moment $\va{x^{-1}}_{\pi}$, 
calculated as $\int_x^{x+0.02}\phi(x) dx$ and normalized to 100\%, 
are shown for CZ and BMS DAs.
%%%%%%%%%%%%%%%%%%%%%%%%%%%%%%%%%%%%%%%%%%%%%%%%%%%%%%%%%%%%%%%%%%%%%%%%%%%%%%%%
%%%%%%%%%%%%%%%%%%%%%%%%%%%%%%%%%%%%%%%%%%%%%%%%%%%%%%%%%%%%%%%%%%%%%%%%%%%%%%%%

\section{LCSR analysis of CLEO data on $F_{\gamma\gamma^*\pi}(Q^2)$ and pion DA}
Why does one need to use Light-Cone SRs (LCSRs) 
in analyzing the experimental data on $\gamma^*(Q)\gamma(q)\to\pi^0$-transition 
form factor?
For $Q^2\gg m_\rho^2$, $q^2\ll m_\rho^2$ 
the QCD factorization is valid only in the leading twist approxi-
\\\noindent
\begin{minipage}{0.49\textwidth}
 mation and the higher twists are of importance~\cite{RR96}.
 The reason is evident: 
 if $q^2\to0$ one needs to take into account interaction 
 of a real photon at long distances 
 of order of $O(1/\sqrt{q^2})$.
 To account for long-distance effects
 in perturbative QCD 
\end{minipage}~~~\begin{minipage}{0.49\textwidth}\vspace*{-4mm}
 $$\includegraphics[width=\textwidth]{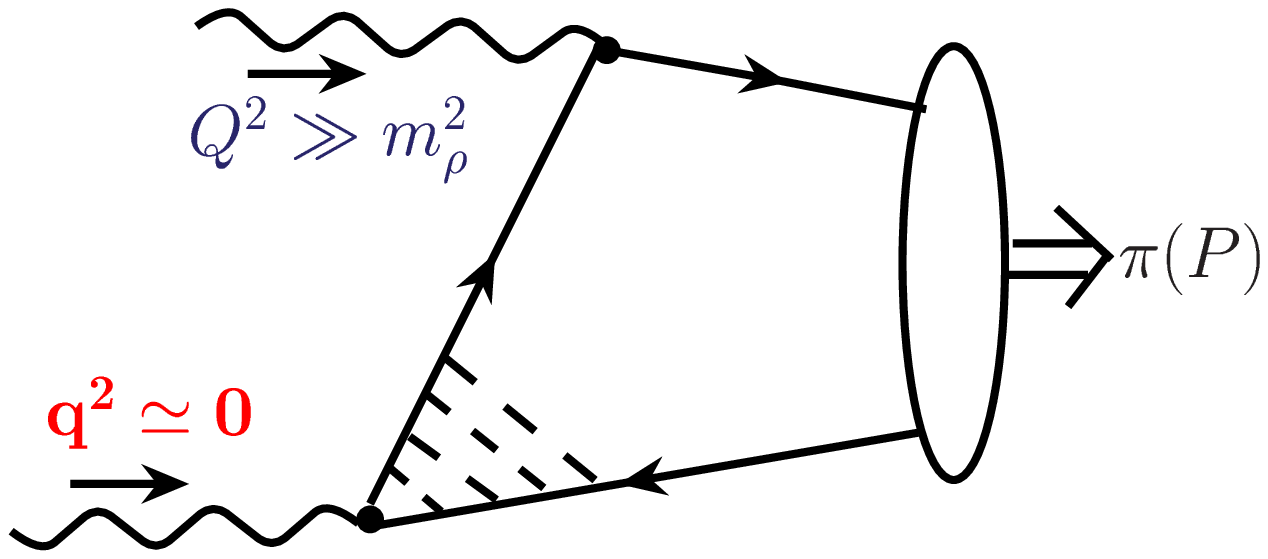}$$
\end{minipage}\vspace*{1pt} 
one needs to introduce the light-cone DA of the real photon.

Instead of doing so,
Khodjamirian~\cite{Kho99} suggested to use the LCSR approach,
which effectively accounts for long-distances effects 
of the real photon using the quark-hadron duality 
in the vector channel and dispersion relation in $q^2$.

We refined the NLO analysis of the CLEO data~\cite{SY99}
by taking into account the following items:\\
(i) an accurate NLO evolution for both {$\varphi(x, Q^2_\text{exp})$} 
    and $\alpha_s(Q^2_\text{exp})$
    with accounting for quark thresholds;\\
(ii) the relation  between the ``nonlocality" scale and 
     the twist-4 magnitude $\delta^2_\text{Tw-4} \approx \lambda_q^2/2$ 
     was used to re-estimate $\delta^2_\text{Tw-4}=0.19 \pm 0.02$ 
     at $\lambda_q^2=0.4$ GeV$^2$;\\
(iii) the possibility to extract constraints on $\va{x^{-1}}_\pi$ 
      from the CLEO data and to compare them with what we have from
      NLC QCD SRs.

\begin{figure}[h]
 \centerline{\includegraphics[width=0.33\textwidth]{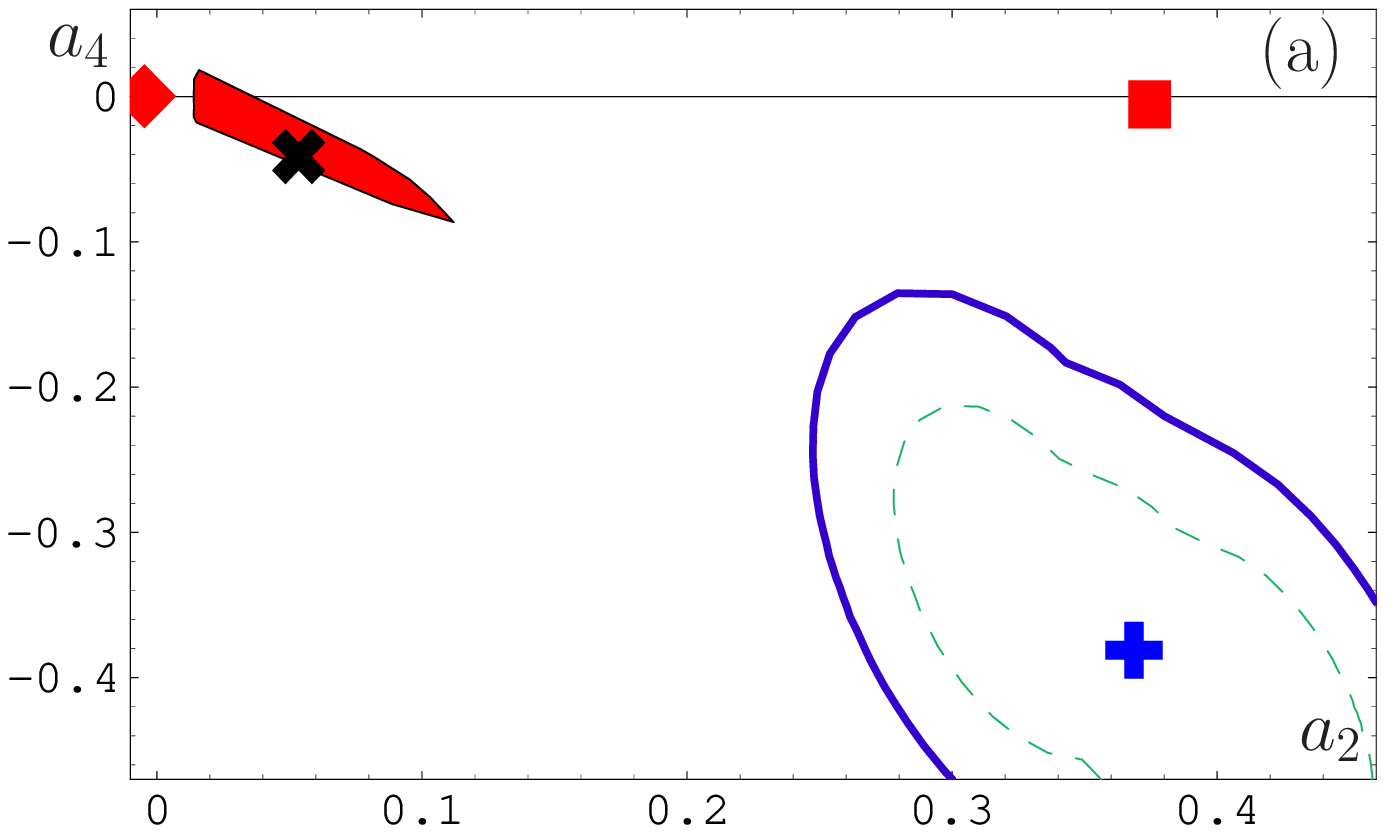}%
            ~\includegraphics[width=0.33\textwidth]{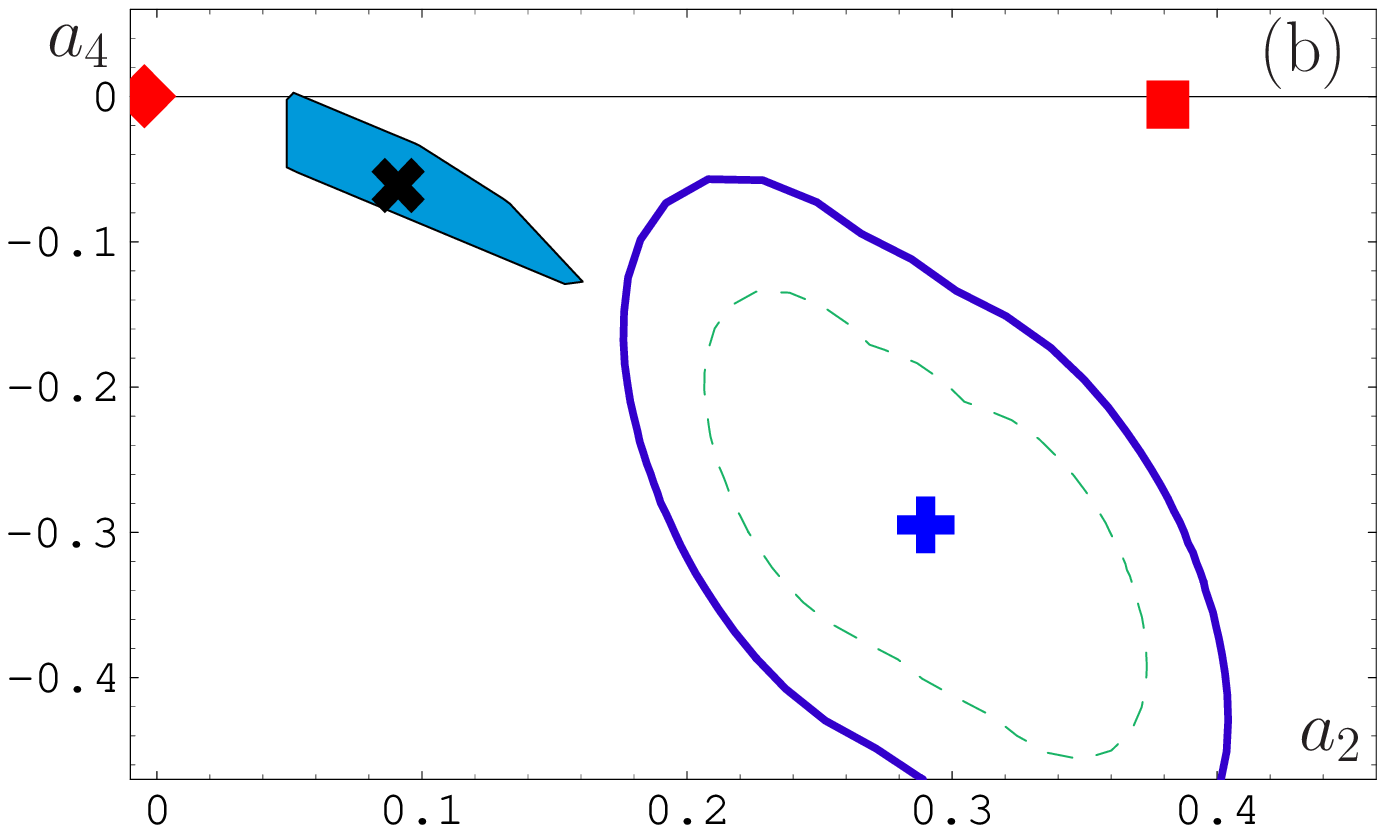}%
            ~\includegraphics[width=0.33\textwidth]{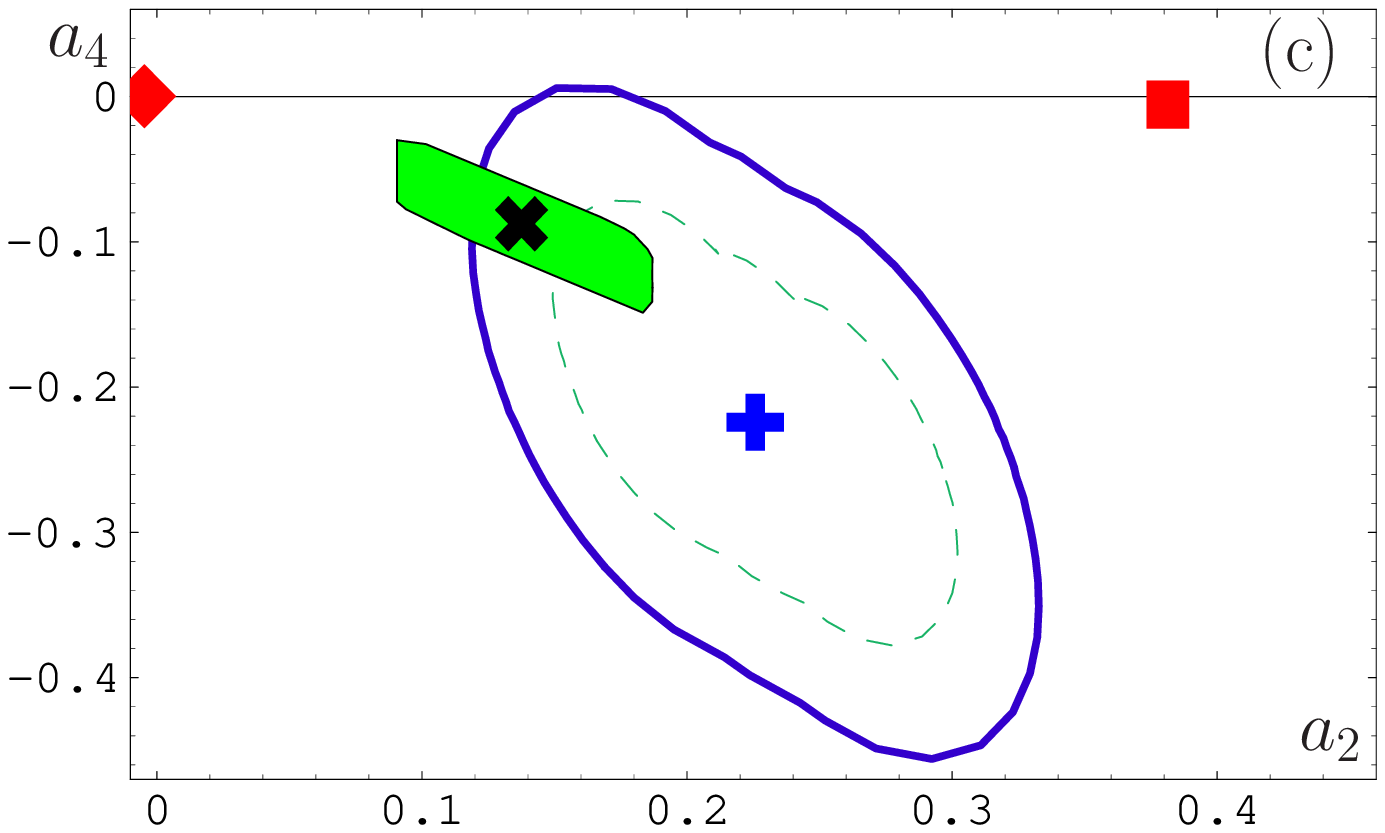}}%
  \caption{\footnotesize
   Three $2\sigma$-contours of the admissible regions
   following from the analysis of the CLEO data for different values
   of $\delta^2$:
   (a) for $\lambda^2_q=0.6~\text{GeV}^2$ and 
    $\delta_\text{Tw-4}^2=(0.29\pm0.03)~\text{GeV}^2$;
   (b) for $\lambda^2_q=0.5~\text{GeV}^2$ and 
    $\delta_\text{Tw-4}^2=(0.235\pm0.025)~\text{GeV}^2$;
   (c) for $\lambda^2_q=0.4~\text{GeV}^2$ and 
    $\delta_\text{Tw-4}^2=(0.19\pm0.02)~\text{GeV}^2$.
   \vspace*{-3mm}\label{fig:cleo.456}}
\end{figure}

The results of our analysis~\cite{BMS02} 
are displayed in Fig.\ \ref{fig:cleo.456}.
%%%%%%%%%%%%%%%%%%%%%%%%%%%%%%%%%%%%%%%%%%%%%%%%%%%%%%%%%%%%%%%%%%%%%%%%%%%%%
%%%%%%%%%%%%%%%%%%%%%%%%%%%%%%%%%%%%%%%%%%%%%%%%%%%%%%%%%%%%%%%%%%%%%%%%%%%%%
Solid lines in all figures enclose the $2\sigma$-contours,
whereas the $1\sigma$-contours are enclosed by dashed lines.
The three slanted and shaded rectangles represent the constraints
on ($a_2,~a_4$) posed by the QCD SRs~\cite{BMS01}
for corresponding values of $\lambda^2_q=0.4,~0.5,~0.6$~GeV$^2$
(from left to right).
All values are evaluated at $\mu^2=2.4$~GeV$^2$ after the NLO evolution.

\begin{figure}[b]\hspace*{-3mm}
 \begin{tabular}{cc}
  \begin{minipage}{0.48\textwidth}\vspace*{-3mm}
   $$\includegraphics[width=\textwidth]{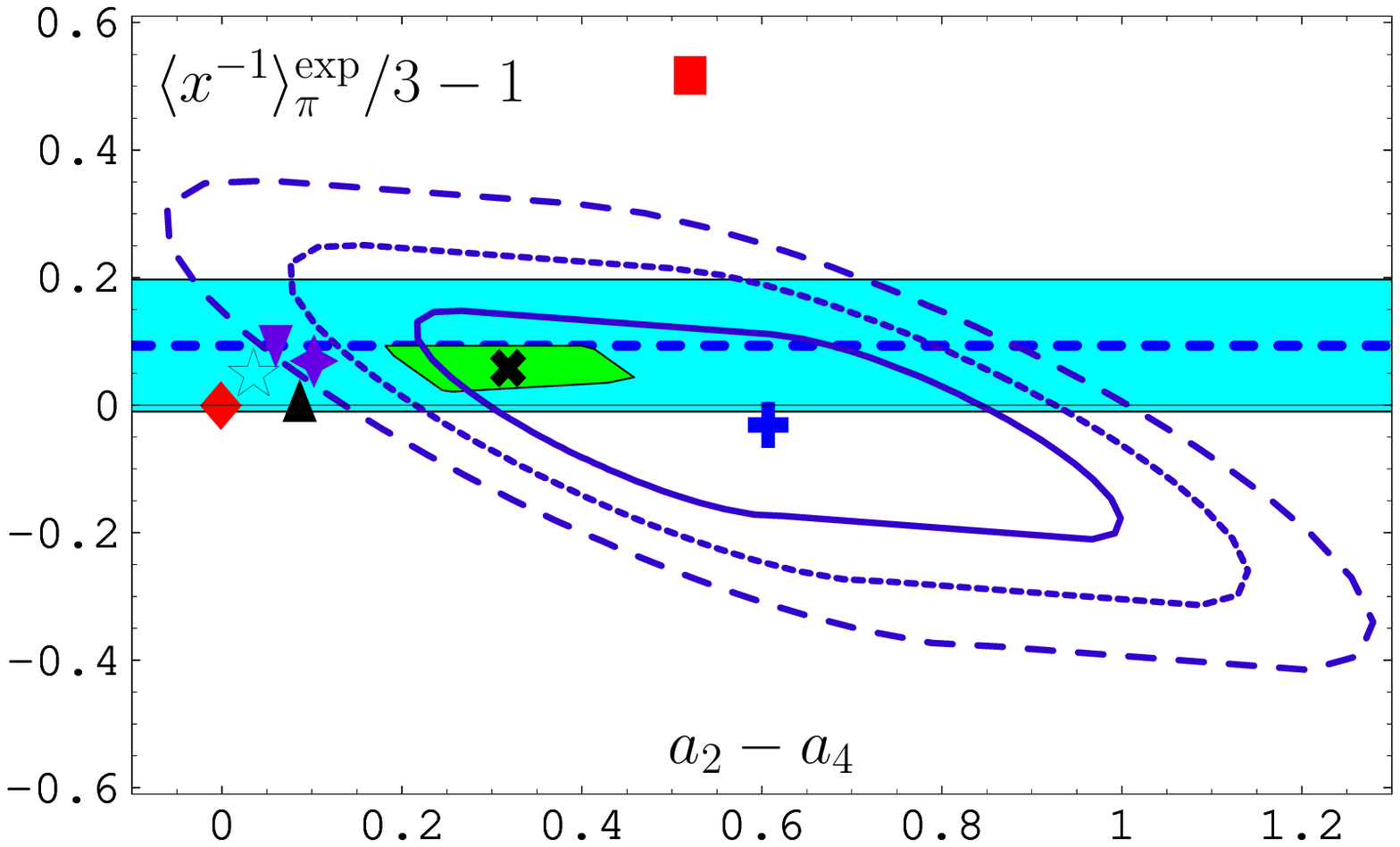}$$
   \end{minipage}&
  \begin{minipage}{0.48\textwidth}\vspace*{-3mm}
   $$\includegraphics[width=\textwidth]{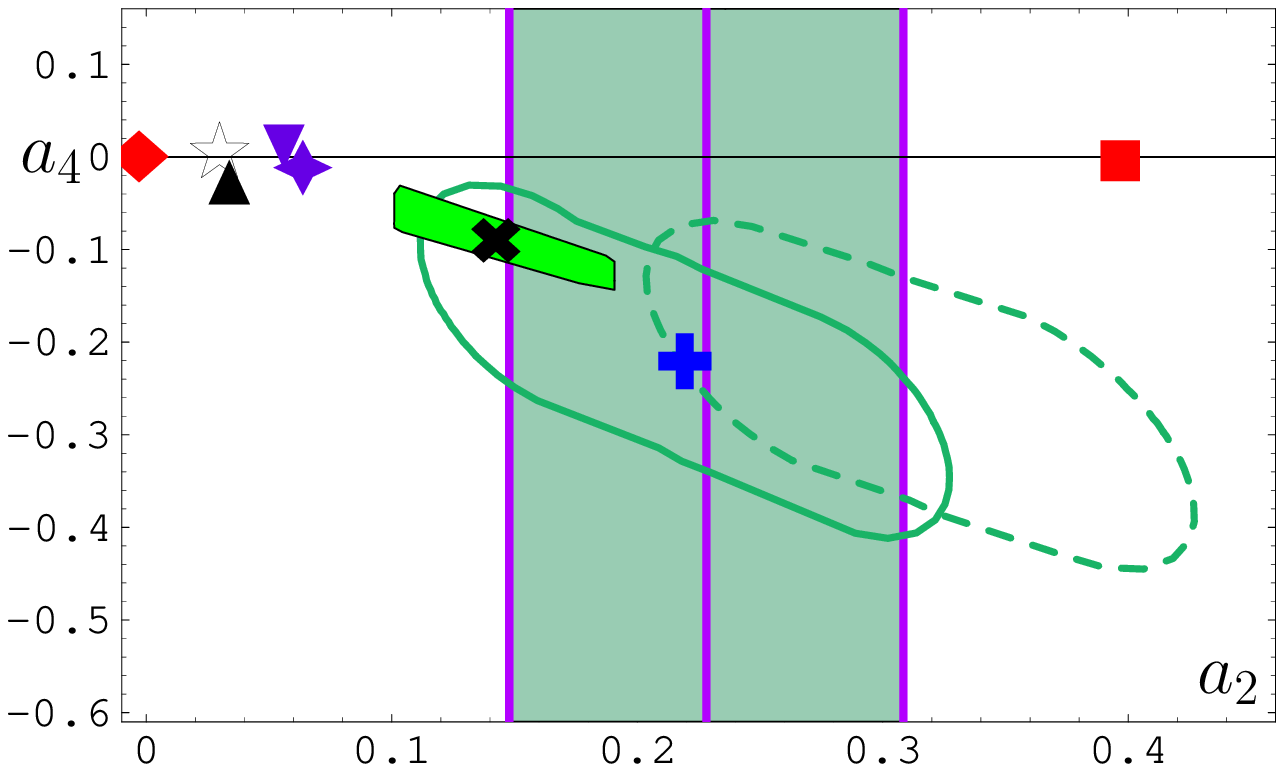}$$
   \end{minipage} 
 \end{tabular}
 \caption{\footnotesize The results of the CLEO data analysis for the pion DA
        parameters 
        ($\displaystyle \langle x^{-1} \rangle^\text{exp}_{\pi}/3-1$,
         evaluated at $\mu^2_0 \approx 1~\text{GeV}^2$, in the \textbf{left panel},
         and $a_2$ and $a_4$, evaluated at $\mu^2_\text{SY}=5.76~\text{GeV}^2$
         -- in the \textbf{right panel}).
        In the \textbf{right panel} the lattice results of \protect\cite{Lat05}
        are shown for comparison as shaded area,
        whereas the renormalon-based $1\sigma$-ellipse of~\protect\cite{BMS05lat} 
        is displayed by the green dashed line.
        \label{fig:Lat.Ren.CLEO}\vspace*{-3mm}}
\end{figure}

We see that the CLEO data definitely prefer the value of
the QCD nonlocality parameter $\lambda_q^2 = 0.4$ GeV$^2$.
We also see in Fig.\ \ref{fig:cleo.456}(c)
(and this conclusion was confirmed even 
 with 20\% uncertainty in twist-4 magnitude, 
 see also Fig.\ \ref{fig:Lat.Ren.CLEO})
that CZ DA ({\red\footnotesize\ding{110}}) is excluded at least at $4\sigma$-level, 
whereas the asymptotic DA ({\red\ding{117}}) --- at $3\sigma$-level.
In the same time our DA (\ding{54}) 
and most of the bunch (the slanted green-shaded rectangle around the symbol \ding{54}) 
are inside 1$\sigma$-domain.
Instanton-based models are all near 3$\sigma$-boundary
and only the Krakow model~\cite{PR01},
denoted in Fig.\ \ref{fig:Lat.Ren.CLEO} 
by symbol {\violet\ding{70}},
is close to 2$\sigma$-boundary.

%%%%%%%%%%%%%%%%%%%%%%%%%%%%%%%%%%%%%%%%%%%%%%%%%%%%%%%%%%%%%%%%%%%%%%%%%%%%%%%%
%%%%%%%%%%%%%%%%%%%%%%%%%%%%%%%%%%%%%%%%%%%%%%%%%%%%%%%%%%%%%%%%%%%%%%%%%%%%%%%%

In the left panel of Fig.\ \ref{fig:Lat.Ren.CLEO} 
we demonstrate the $1\sigma$-, $2\sigma$- and $3\sigma$-contours
(solid, dotted and dashed contours around
 the best-fit point ({\blue\ding{58}})), 
which have been obtained for values of the twist-4 scale parameter 
$\delta_\text{Tw-4}^2=[0.15-0.23]~\text{GeV}^2$.
As one sees from the blue dashed line within the hatched band,
corresponding in this figure to the mean value of 
$\displaystyle \langle x^{-1} \rangle^\text{SR}_{\pi}/3-1$
and its error bars,
the nonlocal QCD sum-rules result with its error bars
appears to be in good agreement with the CLEO-constraints 
on $\displaystyle \langle x^{-1} \rangle^\text{exp}_{\pi}$ 
at the $1\sigma$-level.
Moreover, the estimate $\displaystyle \langle x^{-1} \rangle^\text{SR}_{\pi}$
is close to $\displaystyle \langle x^{-1} \rangle^\text{EM}_{\pi}/3-1=0.24\pm 0.16$,
obtained in the data analysis of the electromagnetic pion form factor
within the framework of a different LCSR method in \cite{BKM00,BK02}.
These three independent estimates are in good agreement to each other,
giving firm support that the CLEO data processing, on one hand, 
and the theoretical calculations, on the other, 
are mutually consistent.

Another possibility, suggested in~\cite{Ag05b}, 
to obtain constraints on the pion DA in the LCSR analysis 
of the CLEO data  -- 
to use for the twist-4 contribution renormalon-based model,
relating it then to parameters $a_2$ and $a_4$ of the pion DA.
Using this method we obtain~\cite{BMS05lat} the renormalon-based constraints 
for the parameters $a_2$ and $a_4$,
shown in the right panel of Fig.\ \ref{fig:Lat.Ren.CLEO} 
in a form of $1\sigma$-ellipses 
(dashed contour).

\textit{New high-precision lattice measurements} of the the pion DA second moment
$\va{\xi^2}_{\pi} = \int_0^1(2x-1)^2\varphi_\pi(x)\,dx$
appeared rather recently~\cite{DelD05,Lat05}.
Both groups extracted from their respective simulations, 
values of $a_2$ at the Schmedding--Yakovlev scale
$\mu^2_\text{SY}$ around $0.24$,
but with different error bars.

It is remarkable that these lattice results are in striking agreement
with the estimates of $a_2$ both from NLC QCD SRs~\cite{BMS01} 
and also from the CLEO-data analyses---based 
on LCSR---\cite{SY99,BMS02}, 
as illustrated in the right panel of Fig.\ \ref{fig:Lat.Ren.CLEO}, 
where the lattice results of~\cite{Lat05}
are shown in the form of a vertical strip, 
containing the central value with associated errors,
being smaller than in~\cite{DelD05}.
Noteworthily, the value of $a_2$ of the displayed lattice measurements
(middle line of the strip) is very close to the CLEO best fit 
in~\cite{BMS02} ({\blue\ding{58}}).

\section{Pion form factor and CEBAF data}
\begin{figure}[t]
 \centerline{\includegraphics[width=0.49\textwidth]{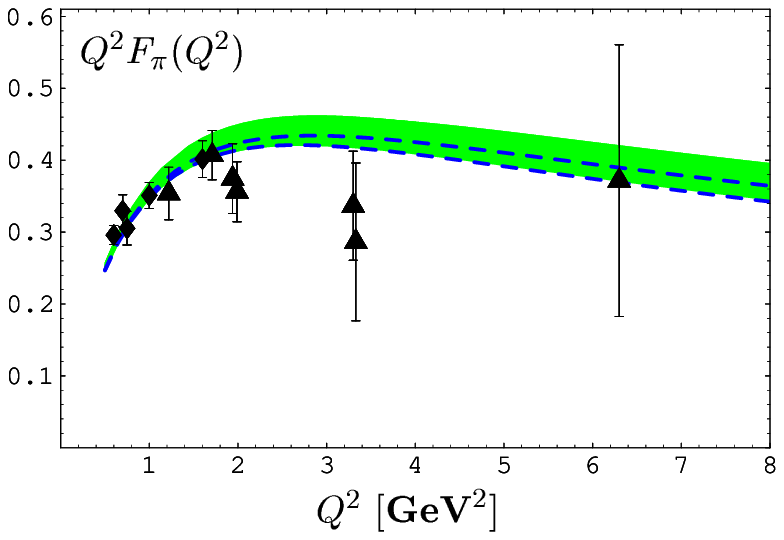}}
  \caption{\footnotesize Predictions for the scaled pion form
    factor calculated with the BMS\ bunch (green strip)
    encompassing nonperturbative uncertainties from nonlocal QCD
    sum rules~\protect{\cite{BMS01}} and renormalization scheme and
    scale ambiguities at the level of the NLO accuracy.
    The dashed lines inside the strip indicate the corresponding
    area of predictions obtained with the asymptotic pion DA.
    The experimental data are taken from \protect{\cite{JLAB00}}
    (diamonds) and \cite{FFPI73}, \cite{FFPI76} (triangles).
    \label{fig:FF.Pi}\vspace*{-1mm}}
\end{figure}

It is worth to mention here the results of our analysis of
the pion electromagnetic form factor
using NLC dictated pion DA and Analytic Perturbative QCD~\cite{BPSS04}.
These results are in excellent agreement
with CEBAF data on pion form factor, 
shown as diamonds in the Fig.\ \ref{fig:FF.Pi},
where the green strip includes both the NLC QCD SRs uncertainties,
generated by our bunch of the allowed pion DA,
and by the scale-setting ambiguities at the NLO level.

%%%%%%%%%%%%%%%%%%%%%%%%%%%%%%%%%%%%%%%%%%%%%%%%%%%%%%%%%%%%%%%%%%%%%%%%%%%%%%%%
%%%%%%%%%%%%%%%%%%%%%%%%%%%%%%%%%%%%%%%%%%%%%%%%%%%%%%%%%%%%%%%%%%%%%%%%%%%%%%%%

From the phenomenological point of view, 
the most interesting result here is 
that the BMS pion DA~\cite{BMS01} 
(out of a ``bunch'' of similar doubly-peaked endpoint-suppressed pion DAs) 
yields to predictions for the electromagnetic form factor 
very close to those obtained with the asymptotic pion DA.
Conversely, we see that a small deviation 
of the prediction
for the pion form factor 
from that obtained with the asymptotic pion DA
does not necessarily imply that the underlying pion DA 
has to be close
to the asymptotic profile.
Much more important is the behavior of the pion DA in the endpoint
region $x\to 0\,, 1$.

\section{Conclusions}
Let me conclude with the following
observations:
\begin{itemize}
\item  NLC QCD SR method for the pion DA gives us 
the admissible bunches of DAs
for each value of $\lambda_q$.

\item NLO LCSR method produces new constraints
on the pion DA parameters ($a_2$ and $a_4$)
in conjunction with the CLEO data.

\item Comparing results of the NLC SRs
with new CLEO constraints allows to fix
the value of QCD vacuum nonlocality: $\lambda_q^2\simeq0.4~\text{GeV}^2$.

\item This bunch of pion DAs agrees well 
with recent lattice data 
and
with JLab F(pi) data on the pion form factor. 
\end{itemize}
I also suggest to the reader to look in
the very interesting discussion 
of the QCD SR approach and its developments
written by one of its creators~\cite{Shif98}.

\section*{Acknowledgments}
This investigation was supported in part 
by the Bogoliubov--Infeld Programme, grant 2006,
by the Heisenberg--Landau Programme, grant 2006, 
and the Russian Foundation for Fundamental Research, grant No.\ 06-02-16215.
   
%%\bibliographystyle{prsty-ab}
%%\bibliography{pion,lambda,nonloc}

\begin{thebibliography}{1}
\bibitem{Rad98}
  A.~V. Radyushkin,
   in {\em Strong Interactions at Low and Intermediate
   Energies: Proceedings of the 13th Annual HUGS AT CEBAF (HUGS 98), 
   26 May--2 Jun 1998, Newport News, Virginia},
   edited by J.~L. Goity (World Scientific, Singapore, 2000), 
   pp.\ 91--150.
  %%CITATION = HEP-PH 0101227;%%.

\bibitem{SVZ}
  M.~A. Shifman, A.~I. Vainshtein, and V.~I. Zakharov,
   Nucl. Phys. \textbf{B147},  385  (1979);
    %%CITATION = NUPHA,B147,385;%%.
   ibid. 448;  
    %%CITATION = NUPHA,B147,448;%%.
   ibid. 519.
    %%CITATION = NUPHA,B147,519;%%.
    
\bibitem{NOSVVZ77}
 V.~A. Novikov {\it et~al.},
  Phys. Rept. \textbf{41},  1  (1978).
   %%CITATION = PRPLC,41,1;%%.

\bibitem{IZ00}
 B.~L. Ioffe and K.~N. Zyablyuk,
  Nucl. Phys. \textbf{A687},  437  (2001);\\
   %%CITATION = HEP-PH 0010089;%%.
 B.~V. Geshkenbein, B.~L. Ioffe, and K.~N. Zyablyuk,
  Phys. Rev. \textbf{D64},  093009  (2001).
   %%CITATION = HEP-PH 0104048;%%. 

\bibitem{Rad77}
 A.~V. Radyushkin,
  \uppercase{D}ubna preprint P2-10717, 1977 [hep-ph/0410276].
    %%CITATION = HEP-PH 0410276;%%.


\bibitem{MR86ev}
 S.~V. Mikhailov and A.~V. Radyushkin,
  Nucl. Phys. \textbf{B273},  297  (1986).
   %%CITATION = NUPHA,B273,297;%%.

\bibitem{KMR86}
 E.~P. Kadantseva, S.~V. Mikhailov, and A.~V. Radyushkin,
  Sov. J. Nucl. Phys. \textbf{44},  326  (1986).
   %%CITATION = YAFIA,44,507;%%.

\bibitem{Mul94}
 D. M{\"u}ller,
  Phys. Rev. \textbf{D49},  2525  (1994);
   %%CITATION = PHRVA,D49,2525;%%.
  ibid. \textbf{D51},  3855  (1995).
   %%CITATION = HEP-PH 9411338;%%.

\bibitem{BS05}
 A.~P. Bakulev and N.~G. Stefanis,
  Nucl. Phys. \textbf{B721},  50  (2005).
   %%CITATION = HEP-PH 0503045;%%.   
   
\bibitem{MR86}
 S.~V. Mikhailov and A.~V. Radyushkin, 
  JETP Lett. \textbf{43},  712  (1986);
   %%CITATION = JTPLA,43,712;%%.
  Sov. J. Nucl. Phys. \textbf{49},  494  (1989);
   %%CITATION = SJNCA,49,494;%%.
  Phys. Rev. \textbf{D45},  1754  (1992).
   %%CITATION = PHRVA,D45,1754;%%.

\bibitem{BM98}
 A.~P. Bakulev and S.~V. Mikhailov,
  Phys. Lett. \textbf{B436},  351  (1998).
   %%CITATION = PHLTA,B436,351;%%.
   
\bibitem{BI82lam}
 V.~M. Belyaev and B.~L. Ioffe,
  ZhETF \textbf{83},  876  (1982).
   %%CITATION = SPHJA,83,876;%%.

\bibitem{OPiv88} 
 A.~A. Ovchinnikov and A.~A. Pivovarov,
  Sov. J. Nucl. Phys. \textbf{48},  721  (1988).
   %%CITATION = SJNCA,48,721;%%.

\bibitem{Piv91}
 A.~A. Pivovarov,
  Bull. Lebedev Phys. Inst. \textbf{5},  1  (1991).
   %%CITATION = SPLRD,5,1;%%.

\bibitem{DDM99}
 M. D'Elia, A. {\uppercase{d}i}~Giacomo, and E. Meggiolaro,
  Phys. Rev. \textbf{D59},  054503  (1999).
   %%CITATION = PHRVA,D59,054503;%%.

\bibitem{BM02}
 A.~P. Bakulev and S.~V. Mikhailov,
  Phys. Rev. \textbf{D65},  114511  (2002).
   %%CITATION = HEP-PH 0203046;%%.

\bibitem{BMS01}
 A.~P. Bakulev, S.~V. Mikhailov, and N.~G. Stefanis,
  Phys. Lett. \textbf{B508},  279  (2001);
   %%CITATION = HEP-PH 0103119;%%.
  in {\em Proceedings of the 36th Rencontres De Moriond 
  On QCD And Hadronic Interactions, 17--24 Mar 2001,
  Les Arcs, France}, edited by J.~T.~T. Van 
  (World Scientific, Singapore, 2002), pp.\ 133--136.
  %%CITATION = HEP-PH 0104290;%%.

\bibitem{CZ82}
 V.~L. Chernyak and A.~R. Zhitnitsky,
  Nucl. Phys. \textbf{B201},  492  (1982);
   %%CITATION = NUPHA,B201,492;%%.
  ibid. \textbf{B214}, 547(E) (1983).   

\bibitem{RR96}
 A.~V. Radyushkin and R. Ruskov,
  Nucl. Phys. \textbf{B481},  625  (1996).
   %%CITATION = HEP-PH 9603408;%%.

\bibitem{Kho99}
 A. Khodjamirian,
  Eur. Phys. J. \textbf{C6},  477  (1999).
   %%CITATION = EPHJA,C6,477;%%.
   
\bibitem{SY99}
 A. Schmedding and O. Yakovlev,
  Phys. Rev. \textbf{D62},  116002  (2000).
   %%CITATION = PHRVA,D62,116002;%%.

\bibitem{BMS02}
 A.~P. Bakulev, S.~V. Mikhailov, and N.~G. Stefanis,
  Phys. Rev. \textbf{D67},  074012  (2003);
   %%CITATION = HEP-PH 0212250;%%.
  Phys. Lett. \textbf{B578},  91  (2004).
   %%CITATION = HEP-PH 0303039;%%.

\bibitem{PR01}
 M. Praszalowicz and A. Rostworowski,
  Phys. Rev. \textbf{D64},  074003  (2001).
   %%CITATION = HEP-PH 0105188;%%.

\bibitem{BKM00}
 V.~M. Braun, A. Khodjamirian, and M. Maul,
  Phys. Rev. \textbf{D61},  073004  (2000).
   %%CITATION = HEP-PH 9907495;%%.

\bibitem{BK02}
 J. Bijnens and A. Khodjamirian,
  Eur. Phys. J. \textbf{C26},  67  (2002).
   %%CITATION = HEP-PH 0206252;%%.

\bibitem{Ag05b}
 S.~S. Agaev,
  Phys. Rev. \textbf{D72},  114010  (2005).
   %%CITATION = HEP-PH 0511192;%%, hep-ph/0511192.

\bibitem{BMS05lat}
 A.~P. Bakulev, S.~V. Mikhailov, and N.~G. Stefanis,
  Phys. Rev. \textbf{D73},  056002  (2006).
   %%CITATION = HEP-PH 0512119;%%.

\bibitem{DelD05}
 L. Del~Debbio,
  Few Body Syst. \textbf{36},  77  (2005).
   %%CITATION = FBSYE,36,77;%%.

\bibitem{Lat05}
 M. G{\"o}ckeler {\it et~al.},
  talk in the Workshop on Light-Cone QCD and Nonperturbative Hadron Physics 
  2005 (LC 2005), Cairns, Queensland, Australia, 7--15 Jul 2005
  [hep-lat/0510089].
   %%CITATION = HEP-LAT 0510089;%%.

\bibitem{BPSS04}
 A.~P. Bakulev \textit{et al.},%%, K. Passek-Kumeri\v{c}ki, W. Schroers, and N.~G. Stefanis,
  Phys. Rev. \textbf{D70},  033014  (2004).
   %%CITATION = HEP-PH 0405062;%%.

\bibitem{JLAB00}
 J. Volmer {\it et~al.},
  Phys. Rev. Lett. \textbf{86},  1713  (2001).
   %%CITATION = NUCL-EX 0010009;%%.

\bibitem{FFPI73}
 C.~N. Brown {\it et~al.},
  Phys. Rev. \textbf{D8},  92  (1973).
   %%CITATION = PHRVA,D8,92;%%.

\bibitem{FFPI76}
 C.~J. Bebek {\it et~al.},
  Phys. Rev. \textbf{D13},  25  (1976).
   %%CITATION = PHRVA,D13,25;%%.
   
\bibitem{Shif98}
 M.~Shifman, 
  ``Snapshots of Hadrons or the Story of How the Vacuum Medium 
    Determines the Properties of the Classical Mesons 
    Which Are Produced, Live and Die in the {QCD} Vacuum'',
   lecture given at the 1997 Yukawa International Seminar on 
   Non-Perturbative QCD-Structure of the QCD Vacuum, 
   Kyoto, December 2--12, 1997 [hep-ph/9802214].
    %%CITATION = HEP-PH 9802214;%%.
   
     
\end{thebibliography}
%%\end{document}

\end{document}